\documentclass[10pt,onecolumn]{IEEEtran}

\usepackage[T1]{fontenc}
\usepackage[utf8]{inputenc}
\usepackage{indentfirst}
\usepackage{amsfonts}
\usepackage{yfonts}
\usepackage{mathrsfs}
\usepackage{multirow}
\usepackage{graphicx}
\usepackage{amsmath}
\usepackage{rotating}
\usepackage{placeins}
\usepackage{amsthm}
\usepackage{amsmath,amssymb}

\def\ie{\emph{i.e.},~}
\def\eg{\emph{e.g.},~} 
\def\st{s.t.~}
\def\wrt{w.r.t.~}
\def\nmp{NMPC$^*$}

\def\calC{{\cal C}}
\def\calS{\mathbf{\cal S}}

\usepackage{tikz}
\usetikzlibrary{automata,patterns,topaths,positioning,shapes.geometric,fit}\usepackage{subfigure}

\newtheorem{proposition}{Proposition}
\newtheorem{theorem}{Theorem}
\newtheorem{lemma}{Lemma}
\newtheorem{corollary}{Corollary}
\newtheorem{definition}{Definition}
\newtheorem{fact}{Fact}

\def\reach{\textit{Reach}}

\def\last{\textit{last}}
\def\trace{\textit{trace}} 
\def\runs{\textit{Runs}}

\def\lang{{\cal L}}

\newcommand{\proj}[1]{\mathbf{proj}_{#1}}

\newcommand{\M}{A}

\newcommand{\R}{\mathcal{R}}

\newcommand{\T}{\mathcal{T}}

\def\untimed{\textit{Untimed}}

\def\weakbisim{\approx_{\fontsize{4}{6}\selectfont \cal W}}

\def\weakcosim{\approx_{\fontsize{4}{6}\selectfont \cal CW}}
\def\weaksim{\sqsubseteq_{\fontsize{4}{6}\selectfont \cal W}}

\def\eqlang{\approx_{\cal L}}
\def\TAldet{\textit{dTA}}
\def\bad{\textit{Bad}}

\newtheorem{example}{Example}
\newtheorem{remark}{Remark}

\def\setR{\mathbb{R}}
\def\setRp{\mathbb{R}_+}
\def\setZ{\mathbb{N}}
\def\setB{\mathbb{B}}
\newcommand{\true}{\textsf{true}} % 
\newcommand{\false}{\textsf{false}}

\makeatletter
\newcommand{\xRightarrow}[2][]{\ext@arrow 0359\Rightarrowfill@{#1}{#2}}
\makeatother
\newcommand{\vect}[1]{\overline{#1}}

\tikzset{initial text={},
    every state/.style={thick,minimum size=.2cm,draw=none},
    secret/.style={minimum size=.2cm,draw=red!50,very thick,fill=red!20,rectangle},
    node distance=2cm,
    on grid,
    auto,
    thick,
    bend angle=65,>=latex}

\renewcommand{\emptyset}{\varnothing}

\begin{document}


\title{Control and Synthesis of Non-Interferent Timed Systems}

\author{Gilles~Benattar,~Franck Cassez,~Didier~Lime~and~Olivier~H.~Roux,
\IEEEcompsocitemizethanks{\IEEEcompsocthanksitem G. Benattar is with ClearSy (Safety Critical Systems Engineering Company) Paris, France.%\protect\\
\IEEEcompsocthanksitem D. Lime and O. H. Roux are with IRCCyN laboratory, LUNAM Universit\'e, Ecole Centrale Nantes,
%1 rue de la No\"{e}, 44 300 Nantes, 
France.%\protect\\
%E-mail: didier.lime@ircyyn.ec--nantes.fr, olivier-h.roux@irccyn.ec-nantes.fr\protect\\
\IEEEcompsocthanksitem F. Cassez is with National ICT Australia, Sydney, Australia.}
}

\maketitle
%\IEEEpeerreviewmaketitle

%\IEEEcompsoctitleabstractindextext{% 
\begin{abstract} 
  In this paper, we focus on the synthesis of secure timed systems
  which are modelled as timed automata.  The security property that
  the system must satisfy is a \emph{non-interference} property.
  Intuitively, non-interference ensures the absence of any causal
  dependency from a high-level domain to a lower-level domain.
  Various notions of non-interference have been defined in the
  literature, and in this paper we focus on \emph{Strong
    Non-deterministic Non-Interference} (SNNI) and two (bi)simulation
  based variants thereof (CSNNI and BSNNI).  We consider 

  \emph{timed} non-interference properties for timed systems specified
  by \emph{timed automata} and we study the two following problems: ($1$)
check whether it is possible to find a sub-system so that it is
non-interferent; if yes ($2$) compute a (largest) sub-system which is
non-interferent.

\end{abstract}

% not want either math or citations to appear in the abstract.

\begin{IEEEkeywords}
 Non-Interference,  Timed Automaton,  Safety Timed Games, Control, Synthesis
\end{IEEEkeywords}
%}

%\newpage
%\tableofcontents
%\newpage

% \FC{TODO list
%   \begin{itemize}
%   \item spell-check \textcolor{red}{DONE}
%   \item questions about properties required for TTS, page 5: add time additivity etc \OR{DONE}
%   \item def of semantics of TA to include \OR{DONE}
%   \item def of synchronized product to include \OR{DONE}
%      Utilise-t-on le produit de TA qqart ?
%   \item Fig. 7: patch it [DONE] - \OR{revennir a l'ancienne figure si reponse = true} \FC{peut etre
%     est-ce plus simple avec les deux transitions ?} \OR{OK}
%   \item Theorem 2: proof should be detailed and updated as Fig 7 changed - \OR{lire reponse} \FC{la preuve devrait
%     etre neanmoins plus detaille car c'est la seule contribution de cette section \FC{DONE but \textcolor{red}{check it !!}}.}
%   \item intro/contribution: say what is new vs what was in FORMATS 2009 [DONE] 
%   \item Section V: on lower bound, say this is an optimal algorith
%     and only completeness is missing [DONE]
% \item Add references to support claim that non-interference is important  ...\OR{[Done]} \FC{Very Good!}
%   \end{itemize}
% }

\section{Introduction}

Modern computing environments allow the use of programs that are sent
or fetched from different sites.  Such programs may deal with secret
information such as private data (of a user) or classified data (of an
organization).  One of the basic concerns in such a context is to
ensure that the programs do not leak sensitive data to a third party,
either maliciously or inadvertently.  This is often called {\em
  secrecy\/}.

In an environment with two parties, {\em information flow analysis\/}
defines secrecy as: ``high-level information never flows into
low-level channels''.
Such a definition is referred to as a {\em non-interference\/}
property, and may capture any causal dependency bet\-ween high-level
and low-level behaviors.

\medskip

We assume that there are two users and the set of actions of the
system $S$ is partitioned into $\Sigma_h$ (high-level actions) and
$\Sigma_l$ (low-level actions).  The non-interference properties we
focus on are strong non-deterministic non-in\-ter\-fe\-rence (SNNI),
co\-si\-mu\-lat\-ion-ba\-sed strong non-deterministic non-interference
(CSNNI) and bisi\-mu\-lation-based strong non-deterministic
non-interference (BSNNI). The \emph{non-interference verification
  problem}, for a given system $S$, consists in checking whether $S$
is non-interferent. It is worth noticing that non-interferent
properties are out of the scope of the common sa\-fe\-ty/li\-ve\-ness
classification of system properties~\cite{focardi01classification}.

There is a large body of works on the use of static analysis
techniques to guarantee information flow policies. A general overview
can be found in~\cite{sabelfeld03languagebased}.  Verification of
information flow security
properties~\cite{focardi01classification,focardi97compositional}
%is a very active domain as it 
can be applied to the analysis of cryptographic protocols where many
uniform and concise characterizations of information flow security
properties ({\em e.g.\/} confidentiality, authentication,
non-repudiation or anonymity) in terms of non-interference have been
proposed. For example, the Needham-Schroeder protocol can be proved
insecure by defining the security property using SNNI~\cite{Needham},
and other examples of the use of non-interference in computer systems and
protocols for checking security properties can be found
%Hence, the property of non-interference is widely used
in~\cite{bossi-JCS-07,barthe-ESOP-07,kammuller-FAC-08,krohn-SP-09}
%\marginpar{\FC{\small we should cite papers using non-interference}} \OR{Done}
%in the formal verification of security in computer systems and
%protocols.

In case a system is not non-interferent,  it is interesting to investigate
how and if it cam be rendered non-interferent.
%If the property is not satisfied, then the behavior of the
%system must be modified in order to ensure its security.

This is the scope of this paper where we consider the problem of
\emph{synthesizing} non-interferent timed systems.
In contrast to verification, the \emph{non-interference synthesis
  problem} assumes the system is \emph{open}, \ie we can restrict the
behaviors of $S$: some events, from a particular set $\Sigma_c
\subseteq \Sigma_l \cup \Sigma_h$, of $S$ can be disabled.
The \emph{non-interference control problem} for a system $S$ asks the
following: ``Is there a controller $C$ s.t. $C(S)$ is
non-interferent?'' where $C(S)$ is ``$S$ controlled by $C$''.
The associated \emph{synthesis problem} asks to compute a witness
controller $C$ when one exists.

\smallskip
%We start by studying the SNNI control problem for timed automata because SNNI is a very interesting  notion of non-interference. Still a
As mentioned earlier, SNNI is expressive enough for example to prove
that the Needham-Schroeder protocol is flawed~\cite{Needham}.
Controller synthesis enables one to find automatically the patch(es)
to apply to make such a protocol secure. The use of dense-time to
model the system clearly gives a more accurate and realistic model for
the system and a potential attacker that can measure time.
%We then focus on more restrictive non-interference properties by studying  CSNNI and BSNNI control problems.

\noindent\textbf{\it \bfseries Related Work.}
In~\cite{vandermeyden-vodca-06} the authors consider the complexity of
many non-interfe\-ren\-ce \emph{verification} problems but synthesis
is not addressed.
In~\cite{DSouzaRS05} an exponential time decision procedure for
checking whether a finite state system satisfies a given Basic
Security Predicate (BSP) is presented but the synthesis problem is not
addressed.
Recently supervisory control for opacity property has been studied
in~\cite{Saboori-CDC-08,cassez-atva-09,cassez-FMSD-12} in the untimed
setting. Opacity is undecidable for timed systems~\cite{cassez-isa-09} and
thus the associated control problem is undecidable as well.
In~\cite{cassez-mmm-07} the controller synthesis problem for
non-interference properties is addressed for untimed systems.
In~\cite{yeddes-TAC-09}, supervisory control to enforce Intransitive
non-interference for three level security systems is proposed in the
untimed setting.

The non-interference synthesis problem for dense-time systems
specified by timed automata was first considered
in~\cite{gardey-secco-05}.  The non-interference property considered
in~\cite{gardey-secco-05} is the \emph{state} non-interference
property, which is less demanding than the one we consider here.
%
%
%This paper is a follow-up of our previous work \cite{cassez-mmm-07}
%about \emph{non-inter\-fe\-rence control problems} for untimed
%systems.  In \cite{cassez-mmm-07}, we assumed that the security
%domains coincided with the controllable and uncontrollable actions:
%high-level actions ($\Sigma_h$) could be disabled ($\Sigma_c
%=\Sigma_h$) whereas low-level actions ($\Sigma_l$) could not.  We
%studied the synthesis problems for SNNI and BSNNI and proved they are
%decidable. In the present paper we extend the previous work in two
%directions: (1) we release the constraint $\Sigma_c =\Sigma_h$ and (2)
%consider the synthesis problem for timed automata.  
This paper extends the results of~\cite{benattar-formats-09}
  about \emph{SNNI control problems} for timed systems:
  Section~\ref{sec-results} addresses the SNNI control problem for
  timed systems and is a detailed presentation of the result
  of~\cite{benattar-formats-09} with proofs of the theorems that were
  unpublished. Sections~\ref{sec-def-NI} and~\ref{sec-verif} are new
  and the latter provides a new result, Theorem~\ref{thm-bcsnnivp}.
  Section~\ref{sec-bcsnni-csp} addresses the CSNNI and BSNNI control
  problems for timed systems and also contains new results:
  Theorems~\ref{thm-snnicpuntimed},~\ref{thm-snnicpfinite},~\ref{thm-snnicpcspdta}
  and Propositions~\ref{prop-CSNNI-nmp} and~\ref{prop-BSNNI-nmp}.
%
  \iffalse .  We extend this work in three directions: (1) we give all
  the proofs of theorems; (2) we consider both verification and
  control problems; (3) we consider SNNI, CSNNI and BSNNI properties
  in timed setting.  \fi
%consider the synthesis problem for timed automata.  

%

%It is also of theoretical interest, because this non-interference
%synthesis problem is \emph{really} more difficult than the
%corresponding verification problem in the sense that we can
%\emph{reduce} the SNNI verification problem to a particular instance
%of the synthesis problem: we just have to take $\Sigma_c = \emptyset$.
%This was not the case for the versions of the synthesis problems
%studied in \cite{cassez-mmm-07}.
%

\smallskip

\noindent\textbf{\it \bfseries Our Contribution.}
In this paper, we first exhibit a class $\TAldet$ of timed automata
for which the SNNI verification problem is decidable. The other main
results are: (1) we prove that deciding whether there is a controller
$C$ for a timed automaton $A$ such that (\st in the following) $C(A)$
is SNNI, is decidable for the previous class $\TAldet$; (2) we reduce
the SNNI controller synthesis problem to solving a sequence of
\emph{safety timed games}; (3) we show that there is not always a most
permissive controller for CSNNI and BSNNI; (4) we prove that the
control problem for CSNNI is decidable for the class $\TAldet$ and
that the CSNNI controller synthesis problem for $\TAldet$ reduces to
the SNNI controller synthesis problem.  We also give the theoretical
complexities of these problems.  \smallskip

\noindent\textbf{\it \bfseries Organization of the paper.}
Section~\ref{sec-prelim} recalls the basics of timed automata, timed
languages and some results on safety timed
games. Section~\ref{sec-def-NI} gives the definition of the
non-interference properties we are interested
in. Section~\ref{sec-verif} addresses the verification of
non-interference properties in the timed
setting. Section~\ref{sec-results} gives the definition of the
non-interference synthesis problem and presents the main result: we
show that there is a largest subsystem which is SNNI and this
subsystem is effectively computable.  Section~\ref{sec-bcsnni-csp}
addresses the control problem and controller synthesis problem for
CSNNI and BSNNI properties.  Finally, we conclude in
Section~\ref{sec-conclu}.

\section{Preliminaries}
\label{sec-prelim}

Let $\setRp$ be the set of non-negative reals and $\setZ$ the set of
integers.  Let $X$ be a finite set of positive real-valued variables
called \emph{clocks}.  A valuation of the variables in $X$ is a
function $X \rightarrow \setRp$, that can be written as a vector of
$\setRp^X$.  We let $\vec{0}_X$ be the valuation \st $\vec{0}_X(x)=0$
for each $x \in X$ and use $\vec{0}$ when $X$ is clear from the
context.  Given a valuation $v$ and $R \subseteq X$, $v[R \mapsto 0]$
is the valuation \st $v[R \mapsto 0](x)=v(x)$ if $x \not\in R$ and $0$
otherwise.  An atomic constraint (over $X$) is of the form $x \bowtie
c$, with $x \in X$, $\bowtie \in \{<,\leq,=,\geq,>\}$ and $c \in
\setZ$.  A (convex) formula is a conjunction of atomic constraints.
$\calC(X)$ is the set of convex formulas.  Given a valuation $v$ (over
$X$) and a formula $\gamma$ over $X$, $\gamma(v)$ is the truth value,
in $\setB=\{\true,\false\}$, of $\gamma$ when each symbol $x$ in
$\gamma$ is replaced by $v(x)$.  If $t \in \setRp$, we let $v+t$ be
the valuation \st $(v+t)(x)=v(x)+t$. We let $|V|$ be the cardinality
of the set $V$.

Let $\Sigma$ be a finite set, $\varepsilon \not\in \Sigma$ and
$\Sigma^\varepsilon=\Sigma\cup\{\varepsilon\}$.  A \emph{timed word}
$w$ over $\Sigma$ is a sequence $w=(\delta_0,a_0)(\delta_1,a_1)\cdots
(\delta_n,a_n)$ s.t.  $(\delta_i,a_i) \in \setRp \times \Sigma $ for
$0 \leq i \leq n$ where $\delta_i$ represents the amount of time
elapsed\footnote{For $i=0$ this is the amount of time since the system
  started.} between $a_{i-1}$ and $a_{i}$.  $T\Sigma^*$ is the set of
timed words over $\Sigma$. We denote by $uv$ the \emph{concatenation}
of two timed words $u$ and $v$.  As usual $\varepsilon$ is also the
empty word
s.t. $(\delta_1,\varepsilon)(\delta_2,a)=(\delta_1+\delta_2,a)$: this
means that language-wise, we can always eliminate the $\varepsilon$
action by taking into account its time interval in the next visible
action.  Given a timed word $w \in T\Sigma^*$ and $L \subseteq \Sigma$
the \emph{projection} of $w$ over $L$ is denoted by $\proj{L}(w)$ and
is defined by $\proj{L}(w) = (\delta_0,b_0)(\delta_1,b_1)\cdots
(\delta_n,b_n)$ with $b_i=a_i$ if $ a_i \in L$ and $b_i = \varepsilon$
otherwise.  The \emph{untimed} projection of $w$, $\untimed(w)$, is
the word $a_0a_1 \cdots a_n$ of $\Sigma^*$.

A \emph{timed language} is a subset of $T\Sigma^*$. Let $L$ be a timed
language, the untimed language of $L$ is $\untimed(L)=\{v \in \Sigma^*
\ | \ \exists w \in L \ \st \ v=\untimed(w) \}$.

\begin{definition}[Timed Transition System (TTS)]
A \emph{timed transition system (TTS)} is a tuple
$\calS=(Q,q_0,\Sigma^\varepsilon,\rightarrow)$ where $Q$ is a set of
states, $q_0$ is the initial state, $\Sigma$ a finite alphabet of
actions, $\rightarrow \subseteq Q \times \Sigma^\varepsilon \cup
\setRp \times Q$ is the transition relation. We use the notation $q
\xrightarrow{e} q'$ if $(q,e,q') \in \rightarrow$.
% and impose that for each $q \in Q, q \xrightarrow{0} q$. %\endef
Moreover, TTS should satisfy the classical time-related conditions
where $d,d' \in \setR_{\geq 0}$: i) time determinism: $({q}
\xrightarrow{d}{q'}) \wedge ({q}\xrightarrow{d}{q''}) \Rightarrow (q'
= q'')$, ii) time additivity: $({q}\xrightarrow{d}{q'}) \wedge ({q'}
\xrightarrow{d'}{q''}) \Rightarrow ({q}\xrightarrow{d+d'}{q''}) $,
iii) null delay: $\forall q: {q}\xrightarrow{0}{q}$, and iv) time
continuity: $({q}\xrightarrow{d}{q'}) \Rightarrow (\forall d'\leq d,
\exists q'', {q}\xrightarrow{d'}{q''})$.
\end{definition}
%\FC{TTS ausually assumes time addivity + 0 + time detrnminism. Why don't we need that ?}

A run $\rho$ of $\calS$ from $q_0$ is a finite sequence of transitions
$\rho= q_0 \xrightarrow{e_1} q_1 \xrightarrow{e_2} \cdots
\xrightarrow{e_n}q_n$ s.t.  $(q_i,e_i,q_{i+1}) \in
\rightarrow$ for $0 \leq i \leq n-1$.  We denote by $\last(\rho)$ the
last state of the sequence \ie the state $q_n$.
We let $\runs(q,\calS)$ be the set of runs from $q$ in $\calS$ and
$\runs(\calS)=\runs(q_0,\calS)$. We write $q \xRightarrow{\
  \varepsilon\ } q'$ if there is a run $q \xrightarrow{\ \varepsilon\
} \cdots \xrightarrow{\ \varepsilon\ } q'$ from $q$ to $q'$ \ie
$\xRightarrow{\ \varepsilon \ } \; \stackrel{\text{def}}{=} \;
(\xrightarrow{\ \varepsilon\ })*$.  Given $a \in \Sigma \cup \setRp$,
we define $\xRightarrow{\ a\ } \; \stackrel{\text{def}}{=} \;
\xRightarrow{\ \varepsilon \ } \xrightarrow{\ a\ } \xRightarrow{\
  \varepsilon \ }$.   
% \OR{Tu ne mets pas le $\varepsilon$ sur le
% $\xRightarrow{\ } $ ? Ca fait double notation non ? J'aurais mis : \ie
% $\xRightarrow{\varepsilon } \; \stackrel{\text{def}}{=} \;
% (\xrightarrow{\ \varepsilon\ })*$.  Given $a \in \Sigma \cup \setRp$,
% we define $\xRightarrow{\ a\ } \; \stackrel{\text{def}}{=} \;
% \xRightarrow{\varepsilon } \xrightarrow{\ a\ }
% \xRightarrow{\varepsilon }$.} 
 We write $q_0 \xrightarrow{\ * \ } q_n$
 if there is a run from $q_0$ to $q_n$.
The set of \emph{reachable} states in $\runs(\calS)$ is
$\reach(\calS)=\{ q \, | \, q_0 \xrightarrow{\ * \ } q \}$.  Each run
can be written in a normal form where delay and discrete transitions
alternate \ie $\rho = q_0 \xrightarrow{\delta_0} \xrightarrow{e_0} q_1
\xrightarrow{\delta_1} \xrightarrow{e_1} \cdots \xrightarrow{\delta_n}
\xrightarrow{e_n} q_{n+1} \xrightarrow{\delta} q'_{n+1} \mathpunct.$
The \emph{trace} of $\rho$ is
$\trace(\rho)=(\delta_0,e_0)(\delta_1,e_1) \cdots (\delta_n,e_n)$.

\smallskip

\begin{definition}[Timed automata (TA)]
  A \emph{timed automaton (TA)} is a tuple $A=(Q,$
  $q_0,X,\Sigma^\varepsilon,E,Inv)$ where: $q_0 \in Q$ is the initial
  location; $X$ is a finite set of positive real-valued clocks;
  $\Sigma^\varepsilon$ is a finite set of actions; $E \subseteq Q
  \times \calC(X) \times \Sigma^\varepsilon \times 2^X \times Q$ is a
  finite set of edges. An edge $(q,\gamma,a,R,q')$ goes from $q$ to
  $q'$, with the guard $\gamma \in \calC(X)$, the action $a$ and the
  reset set $R \subseteq X$; $Inv : Q \rightarrow \calC(X)$ is a
  function that assigns an invariant to any location; we require that
  the atomic formulas of an invariant are of the form $x \bowtie c$
  with $\bowtie \in \{<,\leq\}$.
\end{definition}

A finite (or untimed) automaton $A=(Q,q_0,\Sigma^\varepsilon,E)$ is a
special kind of timed automaton with $X = \emptyset$, and consequently
all the guards and invariants are vacuously true.  A timed automaton
$A$ is \emph{deterministic} if for $(q_1,\gamma,a,R,q_2),
(q_1,\gamma',a,$ $R',q_2') \in E, \gamma \wedge \gamma' \neq \false
\Rightarrow q_2 = q_2' \textit{ and } R=R'$.
We recall that timed automata cannot always be determinized (\ie find
a deterministic TA which accepts the same language as a
non-deterministic one, see~\cite{AlurDill94}), and moreover, checking
whether a timed automaton is determinizable is
undecidable~\cite{finkel05}.
%
%\FC{give the semantics !}
\begin{definition}[Semantics of Timed automata]
The \emph{semantics} of a timed automaton $A=(Q,q_0,X,$
$\Sigma^\varepsilon,E,Inv)$ is the TTS 
%$\calS^A=(Q \times \setRp^X,(q_0,\vec{0}),\Sigma^\varepsilon,\rightarrow)$ defined in the usual way.
$\calS^{\M}=(S,s_0,\Sigma^\varepsilon,\rightarrow)$ with
 $S=Q \times (\setR^+)^X$,
 $s_0 = (q_0,\vec{0})$, and
 $\rightarrow$ defined as follows:
	\begin{flushleft}
	\begin{eqnarray*}
          &&(q,v)\xrightarrow{a} (q',v') \ \text{ iff }\ \exists (q,\gamma,a,R,q') \in E \text{ such that } \left\{ \begin{array}{l} \gamma(v)=\true \\ v' = v[R\mapsto 0] \\ Inv(q')(v')=\true \end{array} \right.
          \\ \\
          &&(q,v)\xrightarrow{\delta } (q,v') \ \text{ iff }\  \left\{ \begin{array}{l} v' = v+\delta \\ \forall \delta'\text{, }  0 \leq \delta' \leq \delta, \\ Inv(q)(v+\delta')=\true \end{array} \right.
    	\end{eqnarray*}	
	\end{flushleft}

\end{definition}

If $s=(q,v)$ is a state of $\calS^A$, we denote by $s + \delta$ the
(only) state reached after $\delta$ time units, \ie $s+
\delta=(q,v+\delta)$.  The sets of runs of $A$ is defined as
$\runs(A)=\runs(\calS^A)$ where $\calS^A$ is the semantics of
$A$.
A timed word $w \in T\Sigma^*$ is \emph{generated} by $A$ if
$w=\trace(\rho)$ for some $\rho \in \runs(A)$. The timed language
generated by $A$, $\lang(A)$, is the set of timed words
generated by $A$.

\begin{definition}[Language equivalence]
Two automata $A$ and $B$ are \emph{language equivalent},
denoted by $A \eqlang B$, if $\lang(A)=\lang(B)$ \ie
they generate the same set of timed words.
\end{definition}

\begin{definition}[Simulation]
  Let $\T_1=(S_1,s_0^1,\Sigma^\varepsilon,\rightarrow_1)$,
  $\T_2=(S_2,s_0^2,$ $\Sigma^\varepsilon,\rightarrow_2)$ be two TTS.
  Let $\R \subseteq S_1 \times S_2$ be a relation \st $\R$ is total
  for $S_2$. $\R$ is a weak simulation of $\T_2$ by $\T_1$ iff:
  \begin{enumerate}
  \item $s_0^1 \R s_0^2$,
  \item $\forall (s,p) \in S_1\times S_2$, such that $s \R p$:
    \begin{itemize}
    \item If $p \xRightarrow{\varepsilon}_2 p'$ then $\exists s'$
      such that $s \xRightarrow{\varepsilon}_1 s'$ and $s' \R p'$,
    \item $\forall a \in \Sigma \cup \setRp$, if $p \xRightarrow{a}_2 p'$ then
      $\exists s'$ such that $s \xRightarrow{a}_1 s'$ and $s' \R p'$.
%    \item $\forall \delta \in \setRp$, if $p \xRightarrow{ \delta}_2
%      p'$ then $\exists s'$ such that $s \xRightarrow{\delta}_1 s'$
%      and $s' \R p'$.
    \end{itemize}
  \end{enumerate}
%  \FC{$\xRightarrow{}$ is not defined} \OR{DONE : dans le paragraphe apres la def des TTS} 
$\T_1$ weakly simulates $\T_2$ if
  there exists a weak simulation $\R$ of $\T_2$ by $\T_1$ and we note
  $\T_1 \weaksim \T_2$. Let $\M_1$ and $\M_2$ be two timed automata,
  we say that $\M_1$ weakly simulates $\M_2$ if the semantics of
  $\M_1$ weakly simulates the semantics of $\M_2$, and we note $\M_1
  \weaksim \M_2$.
\end{definition}

\begin{definition}[Cosimulation]
  Two timed automata $\M_1$ and $\M_2$ are \emph{co-similar} iff $\M_1
  \weaksim \M_2$ and $\M_2 \weaksim \M_1$. We note $\M_1 \weakcosim
  \M_2$
\end{definition}

\begin{definition}[Bisimulation]
  Two timed automata $\M_1$ and $\M_2$ are \emph{bisimilar} iff there
  exists a simulation $\R$ of $\M_2$ by $\M_1$ such that $\R^{-1}$ is
  a weak simulation of $\M_1$ by $\M_2$. We note $\M_1 \weakbisim
  \M_2$.
\end{definition}
Note that when no $\varepsilon$ transition exists, we obtain
\emph{strong} versions of similarity and bisimilarity.

\begin{definition}[Product of timed automata] 
Let $A_1=(Q_1,q_{01},X_1,\Sigma^\varepsilon,E_1,$ $Inv_1)$ and
$A_2=(Q_2,q_{02},X_2,\Sigma^\varepsilon,E_2,Inv_2)$ be two TA with
$X_1 \cap X_2=\emptyset$. Let $\Sigma_a \subseteq \Sigma$. 
The \emph{synchronized product} of $\M_1$ and $\M_2$  \wrt $\Sigma_a$, is
the timed automaton $\M_1 \times_{\Sigma_a} \M_2 =(Q_1 \times Q_2,(q_{01},q_{02}),X_1 \cup X_2,\Sigma^\varepsilon,E,Inv)$ 
where $E$ is defined as follows:

\begin{itemize}
\item $((q_1,q_2),\gamma_1 \wedge \gamma_2,a,R_1 \cup
  R_2,(q'_1,q'_2))\in E$ if $a\in \Sigma_a$, $(q_1,\gamma_1,a,R_1,q'_1) \in
  E_1$ and $(q_2,\gamma_2,a,R_2,q'_2) \in
  E_2$; 
\item $((q_1,q_2),\gamma,a,R,(q'_1,q'_2)) \in E$ if $a \in \Sigma \setminus  \Sigma_a$ and 
$\left\{ \begin{array}{l} 
(q_1,\gamma,a,R,q_1') \in E_1 \text{ and }    q'_{2}=q_{2} \\
  \text{or }  (q_2,\gamma,a,R,q_2') \in E_2 \text{ and } q'_{1}=q_{1} \end{array} \right.$
\end{itemize}
and where $Inv ((q_1,q_2)) = Inv_1(q_1) \wedge Inv_2(q_2)$.
\end{definition}

It means that synchronization occurs only for actions in $\Sigma_a$.
When it is clear from the context we omit the subscript $\Sigma_a$ in
$\times_{\Sigma_a}$.
%\FC{Give def of product ! and the right one :-)}
%\OR{Produit syntaxique. Check :-)}

Moreover, in the sequel we will use two operators on TA: the first one gives an
\emph{abstracted} automaton and simply hides a set of labels $L
\subseteq \Sigma$.  Given a TA $A=(Q,q_0,X,\Sigma^\varepsilon,E,$
$Inv)$ and $L \subseteq \Sigma$ we define the TA
$A/L=(Q,q_0,X,(\Sigma\backslash L)^\varepsilon,E_L,$ $Inv)$ where
$(q,\gamma,a,R,q') \in E_L \iff (q,\gamma,a,R,q') \in E$ for $a \in
\Sigma \backslash L$ and $(q,\gamma,\varepsilon,R,q') \in E_L \iff
(q,\gamma,a,R,q') \in E$ for $a \in L \cup \{\varepsilon\}$.
The \emph{restricted} automaton cuts transitions labeled by the
letters in $L \subseteq \Sigma$: Given a TA $A=(Q,$ $q_0,X,\Sigma,E,Inv)$
and $L \subseteq \Sigma$ we define the TA $A \backslash
L=(Q,q_0,X,\Sigma\backslash L,E_L,Inv)$ where $(q,\gamma,a,R,q') \in
E_L \iff (q,\gamma,a,R,q') \in E$ for $a \in \Sigma \backslash L$.

\medskip

We will also use some results on safety control for timed games which
have been introduced and solved in~\cite{MalerPnueliSifakis95}.

\begin{definition}[Timed Game Automaton (TGA)]
A \emph{Timed Game Automaton (TGA)} $A=(Q,q_0,X,\Sigma,E,Inv)$ is a
timed automaton with its set of actions $\Sigma$ partitioned into
\emph{controllable} ($\Sigma_c$) and \emph{uncontrollable}
($\Sigma_u$) actions.
\end{definition}
Let $A$ be a TGA and $\bad \subseteq Q \times \setRp^X$ be the set of
bad states to avoid. $\bad$ can be written $\cup_{1 \leq i \leq
  k}(\ell_i,Z_i)$, with each $Z_i$ defined as a conjunction of
formulas of $\calC(X)$ and each $\ell_i \in Q$ . The \emph{safety
  control problem} for $(A,\bad)$ is: decide whether there is a
controller to constantly avoid $\bad$. Let $\lambda$ be a fresh
special symbol not in $\Sigma^\varepsilon$ denoting the action ``do
nothing''.

A {\em controller} $C$ for $A$ is a partial function from $\runs(A)$
to $2^{\Sigma_c \cup \{\lambda\}}$.  We require that $\forall \rho \in
\runs(A)$, if $a \in C(\rho) \cap \Sigma_c$ then $\last(\rho)
\xrightarrow{\,a\,} (q',v')$ for some $(q',v')$ and if $\lambda \in
C(\rho)$ then $\last(\rho) \xrightarrow{\,\delta\,} (q',v')$ for some
$\delta >0$.  A controller $C$ is {\em state-based} or
\emph{memoryless} whenever $\forall \rho, \rho' \in \runs(A),
\last(\rho)=\last(\rho')$ implies that $C(\rho)=C(\rho')$.

\begin{remark}\label{rem-sub}
  We assume a controller gives a set of actions that are enabled which
  differs from standard definitions~\cite{MalerPnueliSifakis95} where
  a controller only gives one action. Nevertheless for safety timed
  games, one computes a most permissive controller (if there is one)
  which gives for each state the largest set of actions which are
  safe. It follows that any reasonable (\eg Non-Zeno) sub-controller
  of this most permissive controller avoids the set of bad states.
\end{remark}

$C(A)$ defines ``$A$ supervised/restricted by $C$'' and is inductively
defined by its set of runs:
\begin{itemize}
\item $(q_0,\vec{0}) \in \runs(C(A))$,
\item if $\rho \in \runs(C(A))$ and $\rho \xrightarrow{\ e\ } s' \in
  \runs(A)$, then $\rho \xrightarrow{\ e\ } s' \in \runs(C(A))$ if one
  of the following three conditions holds:
  \begin{enumerate}
  \item $e \in \Sigma_u$,
  \item $e \in \Sigma_c \cap C(\rho)$,
  \item $e \in \setRp \ \text{and}\ \forall \delta \ \st \ 0 \leq
    \delta < e, \last(\rho) \xrightarrow{\ \delta\ } \last(\rho) +
    \delta \wedge \lambda \in C(\rho \xrightarrow{\ \delta \ }
    \last(\rho)+ \delta)$. 
  \end{enumerate}
\end{itemize}
$C(A)$ can also be viewed as a TTS where each state is a run of $A$
and the transitions are given by the previous definition.
$C$ is a \emph{winning} controller for $(A,\bad)$ if $\reach(C(A))
\cap \bad = \emptyset$.
For safety timed games, the results are the
following~\cite{MalerPnueliSifakis95,SouzaMadhusudan02}:
\begin{itemize}
\item it is (EXPTIME-complete to decide whether there is a winning
  controller for a safety game $(A,\bad)$;
\item in case there is one, there is a \emph{most permissive}
  controller which is memoryless on the region graph of the TGA
  $A$. This most permissive controller can be represented by a
  TA. This also means that the set of runs of $C(A)$ is itself the
  semantics of a timed automaton, that can be effectively built from
  $A$.
\end{itemize}

%\begin{itemize}
%  \item $\lang(\M \backslash \Sigma_a) = \lang(\M) \cap T(\Sigma \backslash \Sigma_a)^*$,
%  \item $\lang(\M / \Sigma_a) = \proj{\Sigma \backslash \Sigma_a}(\lang(\M))$.
%\end{itemize}

\section{Formal Definitions of Non-Interference Properties}
\label{sec-def-NI}

In the sequel, we will consider Timed Automata defined on an set of
actions $\Sigma = \Sigma_l \cup \Sigma_h$ with $\Sigma_l \cap \Sigma_h
= \emptyset$, where $\Sigma_h$ are the \emph{high level} actions and
$\Sigma_l$ the \emph{low level} actions.  In order to define the
different classes of non interference properties on an automaton $\M$,
we are going to compare $\M \backslash \Sigma_h$ and $\M / \Sigma_h$
\wrt different criteria.

\subsection{Strong Non-Deterministic Non-Interference (SNNI)}

The property \emph{Strong Non-Deterministic Non-Interference} (SNNI)
% also called \emph{timed SNNI} for timed automata
has been introduced by Focardi and Gorrieri
in~\cite{focardi01classification} as a \emph{trace-based}
generalization of non-interference for concurrent systems. SNNI has
been extended to timed models in~\cite{gardey-secco-05}.
%This property is based on \emph{trace equivalence} and then on languages.

\begin{definition}\label{def-snni}
  A timed automaton $\M$ is \emph{SNNI} iff $\M \backslash \Sigma_h
  \eqlang \M / \Sigma_h$
\end{definition}
Since finite automata are timed automata with no clocks, the
definition also applies to finite automata.

Moreover, as $\lang(\M \backslash \Sigma_h) \subseteq \lang(\M /
\Sigma_h)$, we can give a simple characterization
 of the SNNI property:
\begin{proposition}\label{prop-snni}
  A timed automaton $\M$ is SNNI iff $\lang(\M / \Sigma_h) \subseteq
  \lang(\M \backslash \Sigma_h)$.
\end{proposition}

\begin{example}\label{ex-snni}

  Let us consider the automaton $\M_a$ of figure~\ref{fig-snni1} with
  $\Sigma_h=\{h\}$ and $\Sigma_l=\{\ell\}$. This automaton is not
  SNNI, because $\lang(\M\backslash \Sigma_h)=\varepsilon$ whereas
  $\lang(\M/ \Sigma_h)=\ell$ .  The automaton $\M_b$ is SNNI.

 	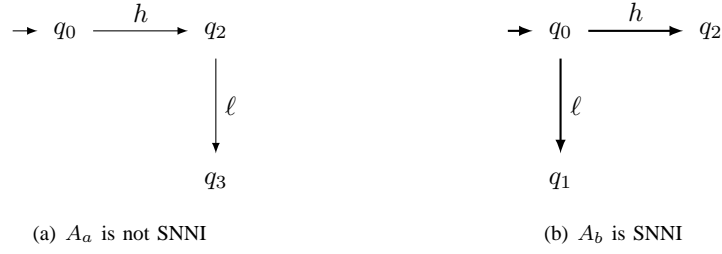
\begin{figure}
	    \begin{center}
		\subfigure[$\M_a$ is not SNNI]{\label{fig-snni1}
		\begin{tikzpicture}[scale=0.7]
			%\gasset{ExtNL=y,NLdist=1,NLangle=90}
			\node[state,initial] (q0) at (0,0) {$q_0$};
			\node[state] (q2) [right of=q0] {$q_2$}; 
			\node[state] (q3) [below of=q2] {$q_3$};   
			%\nodelabel[NLangle=90,NLdist=6](a){$[x_1 \leq 4]$}
			
			\path[->] (q0) edge node {$h$} (q2);
			\path[->] (q2) edge node {$\ell$} (q3);
		\end{tikzpicture}
		}\hspace{3cm}%
		\subfigure[$\M_b$ is SNNI]{\label{fig-snni2}
		\begin{tikzpicture}[node distance=2cm,thick,initial text=,auto,scale=0.7]
			%\gasset{ExtNL=y,NLdist=1,NLangle=90}
			\node[state,initial] (q0) at (0,0) {$q_0$};
			\node[state] (q1) [below of=q0] {$q_1$};
			\node[state] (q2) [right of=q0] {$q_2$};  
			%\nodelabel[NLangle=90,NLdist=6](a){$[x_1 \leq 4]$}

			\path[->] (q0) edge node {$\ell$} (q1);
			\path[->] (q0) edge node {$h$} (q2);
		\end{tikzpicture}
		}
		\caption{Examples for the SNNI property}
		\label{fig-snni12}
	  \end{center}
	\end{figure}
\end{example}

As demonstrated by the following examples~\ref{ex-snnitemp}
and~\ref{ex-snnitemp2}, a timed automaton $\M$ can be non SNNI whereas
its untimed underlying automaton is SNNI and $\M$ can be SNNI whereas
its untimed underlying automaton is not.

\begin{example}\label{ex-snnitemp}
  Let us consider the timed automaton $\M_g$ of
  figure~\ref{fig-snnitemp1}, with $\Sigma_h=\{h\}$ and
  $\Sigma_l=\{\ell\}$. It is not SNNI since $(2.5,\ell)$ is accepted
  by $\M_g /\Sigma_h $ but not by $\M_g \backslash \Sigma_h $. Its
  untimed underlying automaton $\M_h$ is SNNI since
  $\lang(\M_h\backslash \Sigma_h)=\{\ell\}=\lang(\M_h/\Sigma_h)$.
	
 	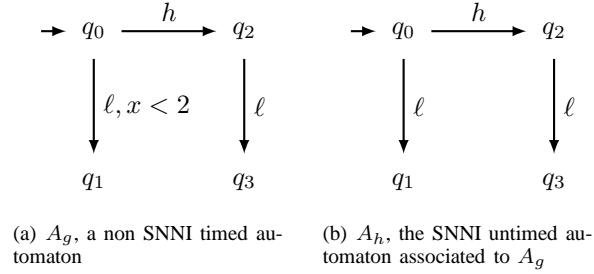
\begin{figure}
	    \begin{center}
		\subfigure[$\M_g$, a non SNNI timed automaton]{\label{fig-snnitemp1}
		\begin{tikzpicture}[node distance=2cm,thick,initial text=,auto,scale=0.7]
			%\gasset{ExtNL=y,NLdist=1,NLangle=90}
			\node[state,initial] (q0) at (0,0) {$q_0$};
			\node[state] (q1) [below of=q0] {$q_1$};
			\node[state] (q2) [right of=q0] {$q_2$};
			\node[state] (q3) [below of=q2] {$q_3$};   
			%\nodelabel[NLangle=90,NLdist=6](a){$[x_1 \leq 4]$}

			\path[->] (q0) edge node {$\ell,x<2$} (q1);
			\path[->] (q0) edge node {$h$} (q2);
			\path[->] (q2) edge node {$\ell$} (q3);
		\end{tikzpicture}
		}\hspace{15pt}%
		\subfigure[$\M_h$, the SNNI untimed automaton associated to $\M_g$]{\label{fig-snnitemp2}
		\begin{tikzpicture}[node distance=2cm,thick,initial text=,auto,scale=0.7]
			%\gasset{ExtNL=y,NLdist=1,NLangle=90}
			\node[state,initial] (q0) at (0,0) {$q_0$};
			\node[state] (q1) [below of=q0] {$q_1$};
			\node[state] (q2) [right of=q0] {$q_2$};
			\node[state] (q3) [below of=q2] {$q_3$};   
			%\nodelabel[NLangle=90,NLdist=6](a){$[x_1 \leq 4]$}

			\path[->] (q0) edge node {$\ell$} (q1);
			\path[->] (q0) edge node {$h$} (q2);
			\path[->] (q2) edge node {$\ell$} (q3);
		\end{tikzpicture}
		}
		\caption{A non SNNI timed automaton and its untimed underlying automaton which is SNNI}
		\label{fig-snnitemp12}
	  \end{center}
	\end{figure}
\end{example}

\begin{example}\label{ex-snnitemp2}
  Let us consider the timed automaton $\M_j$ of
  figure~\ref{fig-snnitemp3}, with $\Sigma_h=\{h\}$ et
  $\Sigma_l=\{\ell_1,\ell_2\}$. It is SNNI, since $\lang(\M_j
  \backslash \Sigma_h)=\lang(\M_j /\Sigma_h)$.  Its untimed underlying
  automaton $\M_k$ is not SNNI since $\ell_1 \cdot \ell2$ is accepted
  by $\M_k /\Sigma_h $ but not by $\M_k \backslash \Sigma_h $.

 	\begin{figure}
	    \begin{center}
		\subfigure[$\M_j$, a SNNI timed automaton]{\label{fig-snnitemp3}
		\begin{tikzpicture}[node distance=2cm,thick,initial text=,auto,scale=0.7]
			%\gasset{ExtNL=y,NLdist=1,NLangle=90}
			\node[state,initial] (q0) at (0,0) {$q_0$};
			\node[state] (q1) [below of=q0] {$q_1$};
			\node[state] (q2) [right of=q0,,shift={(2cm,0cm)}] {$q_3$};
			\node[state] (q3) [below of=q2] {$q_4$};
			\node[state] (q4) [below of=q1] {$q_2$};  
			\node[state] (q5) [below of=q3] {$q_5$};     
			%\nodelabel[NLangle=90,NLdist=6](a){$[x_1 \leq 4]$}

			\path[->] (q0) edge [left] node {$\ell_1,x>2$} (q1);
			\path[->] (q1) edge [left] node {$\ell_1$} (q4);
			\path[->] (q0) edge node {$h$} (q2);
			\path[->] (q2) edge node {$\ell_1, x>2$} (q3);  
			\path[->] (q3) edge [bend left,right] node {$\ell_1$} (q5);
			\path[->] (q3) edge [bend right,left] node {$\ell_2, x<2$} (q5);
		\end{tikzpicture}
		}\hspace{15pt}%
		\subfigure[$\M_k$, the non SNNI untimed automaton associated to $\M_j$]{\label{fig-snnitemp4}
			\begin{tikzpicture}[node distance=2cm,thick,initial text=,auto,scale=0.7]
			%\gasset{ExtNL=y,NLdist=1,NLangle=90}
			\node[state,initial] (q0) at (0,0) {$q_0$};
			\node[state] (q1) [below of=q0] {$q_1$};
			\node[state] (q2) [right of=q0,,shift={(2cm,0cm)}] {$q_3$};
			\node[state] (q3) [below of=q2] {$q_4$};
			\node[state] (q4) [below of=q1] {$q_2$};  
			\node[state] (q5) [below of=q3] {$q_5$};     
			%\nodelabel[NLangle=90,NLdist=6](a){$[x_1 \leq 4]$}

			\path[->] (q0) edge [left] node {$\ell_1$} (q1);
			\path[->] (q1) edge [left] node {$\ell_1$} (q4);
			\path[->] (q0) edge node {$h$} (q2);
			\path[->] (q2) edge node {$\ell_1$} (q3);  
			\path[->] (q3) edge [bend left,right] node {$\ell_1$} (q5);
			\path[->] (q3) edge [bend right,left] node {$\ell_2$} (q5);
		\end{tikzpicture}
		}
		\caption{A SNNI timed automaton and its untimed underlying automaton which is non SNNI.}
		\label{fig-snnitemp34}
	  \end{center}
	\end{figure}
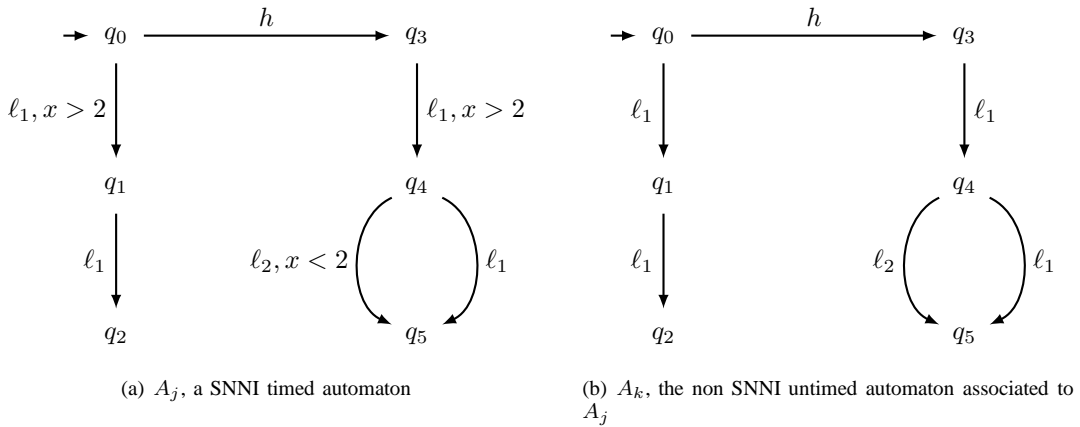
\end{example}

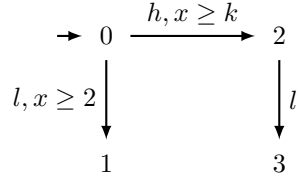
\begin{figure}  \centering
\begin{tikzpicture}[thick,node distance=1.7cm and 2.3cm,initial text=,auto,scale=0.7]% 
    \node[state,initial] (a) {$0$};  %FFF
    \node[state] (b) [below=of a] {$1$};
    \node[state] (c) [right=of a] {$2$};
    \node[state] (d) [below=of c] {$3$};
    \path[->] (a)  edge [swap] node  {$l,x \geq 2$} (b)
                    edge  node  {$h,x \geq k$} (c)
              (c) edge  node  {$l$} (d);
  \end{tikzpicture}
\caption{Automaton $A(k)$}
\label{fig-snni-ex1}
%
%\end{floatingfigure}
\end{figure}

\begin{example}[SNNI]  Figure~\ref{fig-snni-ex1} gives
  examples of systems $A(k)$ which are SNNI and not SNNI depending on
  the value of integer $k$. The high-level actions are
  $\Sigma_h=\{h\}$ and the low-level actions are
  $\Sigma_l=\{l\}$. $(\delta,l)$ with $1 \leq \delta < 2$ is a trace
  of $A(1)/\Sigma_h$ but not of $A(1) \backslash \Sigma_h$ and so,
  $A(1)$ is not SNNI.  $A(2)$ is SNNI as we can see that
  $A(2)/\Sigma_h \eqlang A(2) \backslash \Sigma_h$. 
  %Note that $A(k)$  without the clock constraints, then it is SNNI. 
\end{example}

Finally since SNNI is based on language equivalence, we have the
following lemma:
\begin{lemma}\label{lem-snnieqlang}
  If $\M' \eqlang \M$, then $\M$ is SNNI $\Leftrightarrow$ $\M'$ is
  SNNI.
\end{lemma}

\begin{IEEEproof}
  First $\lang(\M / \Sigma_h)=
  \proj{\Sigma_l}(\lang(\M))=\proj{\Sigma_l}(\lang(\M'))=\lang(\M' /
  \Sigma_h)$.  Second, $\lang(\M \backslash
  \Sigma_h)=\lang(\M) \cap T\Sigma_l^*=\lang(\M') \cap T\Sigma_l^*=\lang(\M' \backslash
  \Sigma_h)$.
% 
%  $\Leftrightarrow$ $(w \in \lang(\M') \cap T\Sigma_l^*)$
%  $\Leftrightarrow$ $w \in \lang(\M' \backslash \Sigma_h)$.  Then we
%  have $\lang(\M \backslash \Sigma_h) = \lang(\M' \backslash
%  \Sigma_h)$.
 
%
%Therefore, if $\M$ is SNNI, $\lang(\M' / \Sigma_h)= \lang(\M / \Sigma_h)= \lang(\M \backslash \Sigma_h)=\lang(\M' \backslash \Sigma_h)$ and then $\M'$ is SNNI and reciprocally.
\end{IEEEproof}

\subsection{Cosimulation Strong Non-Deterministic Non-Interference (CSNNI)}

The \emph{Cosimulation Strong Non-Deterministic Non-Interference}
(CSNNI) property has been introduced in~\cite{gardey-secco-05}, and is
based on \emph{cosimulation}. % (\FC{cosim $\weakcosim$ has not been  defined}).

\begin{definition}
  A timed automaton $\M$ is \emph{CSNNI} iff $\M \backslash \Sigma_h
  \weakcosim \M / \Sigma_h$.
\end{definition}

Since $\M / \Sigma_h \weaksim \M \backslash \Sigma_h$, we can give a
simple characterization of CSNNI:

\begin{proposition}\label{def-snni2}
  A timed Automaton $\M$ is CSNNI iff $\M \backslash \Sigma_h \weaksim
  \M / \Sigma_h$.
\end{proposition}
%\FC{$\weaksim$ not defined} 
By restricting the class of timed automata
considered, we obtain the following result.

\begin{example}\label{ex-snnipascsnni}
 
  Let us consider the automaton $\M_c$ of
  figure~\ref{fig-snnipascsnni1} with $\Sigma_h=\{h\}$ and
  $\Sigma_l=\{\ell_1, \ell_2, \ell_3\}$. $\M_c$ is SNNI but is not
  CSNNI, because no state of $\M_c \backslash \Sigma_l$ can simulate
  the state $q_6$. The automaton $\M_d$ of
  figure~\ref{fig-snnipascsnni1} is CSNNI.  The state $q_1$ of $\M_d
  \backslash \Sigma_l$ simulates the states $q_5$ and $q_6$.

 	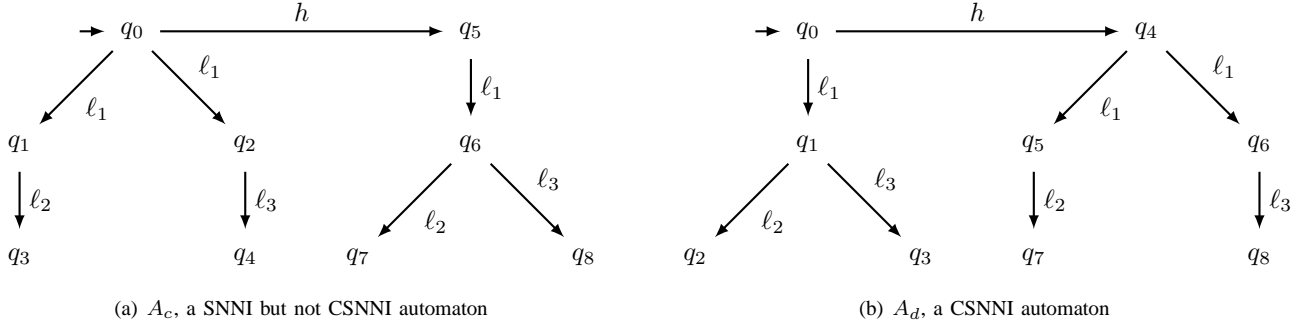
\begin{figure}
	    \begin{center}
	    
		\subfigure[$\M_c$, a SNNI but not CSNNI automaton]{\label{fig-snnipascsnni1}
		\begin{tikzpicture}[node distance=1.5cm,thick,initial text=,auto,scale=0.7]
			%\gasset{ExtNL=y,NLdist=1,NLangle=90}
			\node[state,initial] (q0) at (0,0) {$q_0$};
			\node(q) [below of=q0] {};
			\node[state] (q1) [left of=q] {$q_1$};
			\node[state] (q2) [right of=q] {$q_2$};
			\node[state] (q3) [below of=q1] {$q_3$};
			\node[state] (q4) [below of=q2] {$q_4$};
			\node(qq) [right of=q0] {};
			\node(qqq) [right of=qq] {};
			\node[state] (q5) [right of=qqq] {$q_5$};
			\node[state] (q6) [below of=q5] {$q_6$}; 
			\node(qp) [below of=q6] {}; 
			\node[state] (q7) [left of=qp] {$q_7$};  
			\node[state] (q8) [right of=qp]  {$q_8$};

			%\nodelabel[NLangle=90,NLdist=6](a){$[x_1 \leq 4]$}

			\path[->] (q0) edge node {$\ell_1$} (q1);
			\path[->] (q0) edge node {$\ell_1$} (q2);
			\path[->] (q1) edge node {$\ell_2$} (q3);
			\path[->] (q2) edge node {$\ell_3$} (q4);
			\path[->] (q0) edge node {$h$} (q5);
			\path[->] (q5) edge node {$\ell_1$} (q6);
			\path[->] (q6) edge node {$\ell_2$} (q7);
			\path[->] (q6) edge node {$\ell_3$} (q8);
		\end{tikzpicture}
		}\hspace{15pt}%
		\subfigure[$\M_d$, a CSNNI automaton]{\label{fig-snnipascsnni2}
		\begin{tikzpicture}[node distance=1.5cm,thick,initial text=,auto,scale=0.7]
			%\gasset{ExtNL=y,NLdist=1,NLangle=90}
			\node[state,initial] (q0) at (0,0) {$q_0$};
			\node[state] (q1) [below of=q0] {$q_1$};
			\node(q) [below of=q1] {};
			\node[state] (q2) [left of=q] {$q_2$};
			\node[state] (q3) [right of=q] {$q_3$};
			\node(qq) [right of=q0] {};
			\node(qqq) [right of=qq] {};
			\node[state] (q4) [right of=qqq] {$q_4$};
			\node(qp) [below of=q4] {};
			\node[state] (q5) [left of=qp] {$q_5$};  
			\node[state] (q6) [right of=qp]  {$q_6$};
			\node[state] (q7) [below of=q5] {$q_7$};
 			\node[state] (q8) [below of=q6] {$q_8$};

			%\nodelabel[NLangle=90,NLdist=6](a){$[x_1 \leq 4]$}

			\path[->] (q0) edge node {$\ell_1$} (q1);
			\path[->] (q1) edge node {$\ell_2$} (q2);
			\path[->] (q1) edge node {$\ell_3$} (q3);
			\path[->] (q0) edge node {$h$} (q4);
			\path[->] (q4) edge node {$\ell_1$} (q5);
			\path[->] (q4) edge node {$\ell_1$} (q6);
			\path[->] (q5) edge node {$\ell_2$} (q7);
			\path[->] (q6) edge node {$\ell_3$} (q8);
		\end{tikzpicture}
		}
		\caption{CSNNI is stronger than SNNI}
		\label{fig-snnipascsnni12}
	  \end{center}
	\end{figure}
\end{example} 

We complete this subsection by comparing SNNI and CSNNI.  Given
  two timed automata $A_1,A_2$, $A_1 \weaksim A_2$ implies $\lang(A_2)
  \subseteq \lang(A_1)$.  CSNNI is thus stronger than SNNI as for each
  timed automaton $\M$, $\M \backslash \Sigma_h \weaksim \M /
  \Sigma_h$ implies $\lang( \M / \Sigma_h) \subseteq \lang(\M
  \backslash \Sigma_h)$.

The converse holds when $\M\backslash \Sigma_h$ is deterministic:
\begin{lemma}\label{lem-csnnisnnidet}
  If $\M\backslash \Sigma_h$ is deterministic, then $\M$ is SNNI
  implies $\M$ is CSNNI.
\end{lemma}
 
\begin{IEEEproof}
  As emphasized before, given two timed automata $A_1,A_2$, $A_1
  \weaksim A_2$ implies $\lang(A_2) \subseteq \lang(A_1)$.  If $A_1$
  is deterministic, then $\lang(A_2) \subseteq \lang(A_1)$ implies
  $A_1 \weaksim A_2$.  To obtain the result it suffices to take $A_1=\M
  \backslash \Sigma_h$ and $A_2=\M / \Sigma_h$.

%   It is known that $\forall \M_1, \M_2$ , if $\M_2$ is deterministic,
%   then $\M_2 \weaksim \M_1 \Leftrightarrow \lang(\M_1) \subseteq
%   \lang(\M_2)$.  Let us consider $\M_2=\M \backslash \Sigma_h$ et
%   $\M_1=\M/\Sigma_h$, we have $\lang(\M \backslash \Sigma_h) \subseteq
%   \lang(\M/\Sigma_h)$ and $\M/\Sigma_h \weaksim \M \backslash
%   \Sigma_h$.  Thus, if $\M \backslash \Sigma_h$ is deterministic, then
%   $\M \backslash \Sigma_h \weaksim \M/\Sigma_h \Leftrightarrow
%   \lang(\M/\Sigma_h) = \lang(\M \backslash \Sigma_h)$
\end{IEEEproof}

%\subsection{BSNNI}
\subsection{Bisimulation Strong Non-Deterministic Non-Interference (BSNNI)}

The \emph{Bisimulation Strong Non-Deterministic Non-Interference}
(BSNNI) property has been introduced in~\cite{focardi01classification}
and is based on bisimulation.

\begin{definition}
  A timed automaton $\M$ is BSNNI iff $\M \backslash \Sigma_h
  \weakbisim \M / \Sigma_h$
\end{definition}
%\FC{$\weakbisim$ not defined}

The automaton $\M_f$ of figure~\ref{fig-csnnipasbsnni2} is BSNNI.
Bisimulation is stronger than cosimulation and we have for all timed
automaton $\M$, if $\M$ is BSNNI then $\M$ is CSNNI (and thus $\M$ is
SNNI).

As the following example demonstrates, there exists an automaton which
is CSNNI and not BSNNI.

\begin{example}\label{ex-csnnipasbsnni}

  Let us consider the automaton $\M_e$ of
  figure~\ref{fig-csnnipasbsnni1} with $\Sigma_h=\{h\}$ et
  $\Sigma_l=\{\ell\}$. This automaton is deterministic and SNNI, and
  therefore by lemma~\ref{lem-csnnisnnidet}, it is CSNNI. However, it
  is not BSNNI, since the state $q_2$ of $\M_e \backslash \Sigma_h$
  has no bisimilar state in $\M_e \backslash \Sigma_h$.

 	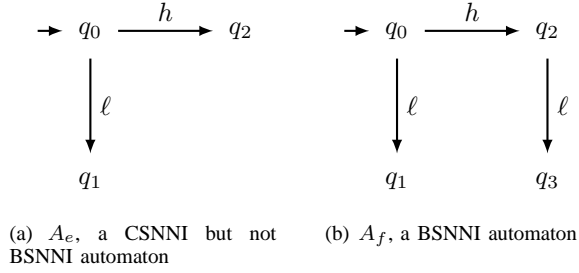
\begin{figure}
	    \begin{center}
		\subfigure[$\M_e$, a CSNNI but not BSNNI automaton]{\label{fig-csnnipasbsnni1}
		\begin{tikzpicture}[node distance=2cm,thick,initial text=,auto,scale=0.7]
			%\gasset{ExtNL=y,NLdist=1,NLangle=90}
			\node[state,initial] (q0) at (0,0) {$q_0$};
			\node[state] (q1) [below of=q0] {$q_1$};
			\node[state] (q2) [right of=q0] {$q_2$};   
			%\nodelabel[NLangle=90,NLdist=6](a){$[x_1 \leq 4]$}

			\path[->] (q0) edge node {$\ell$} (q1);
			\path[->] (q0) edge node {$h$} (q2);
		\end{tikzpicture}
		}\hspace{15pt}%
		\subfigure[$\M_f$, a BSNNI automaton]{\label{fig-csnnipasbsnni2}
		\begin{tikzpicture}[node distance=2cm,thick,initial text=,auto,scale=0.7]
			%\gasset{ExtNL=y,NLdist=1,NLangle=90}
			\node[state,initial] (q0) at (0,0) {$q_0$};
			\node[state] (q1) [below of=q0] {$q_1$};
			\node[state] (q2) [right of=q0] {$q_2$};  
			\node[state] (q3) [below of=q2] {$q_3$}; 
			%\nodelabel[NLangle=90,NLdist=6](a){$[x_1 \leq 4]$}

			\path[->] (q0) edge node {$\ell$} (q1);
			\path[->] (q0) edge node {$h$} (q2);
			\path[->] (q2) edge node {$\ell$} (q3);
		\end{tikzpicture}
		}
		\caption{BSNNI is stronger than CSNNI}
		\label{fig-csnnipasbsnni12}
	    \end{center}
	\end{figure}
\end{example}

\section{Verification of Non-Interference Properties for Timed Automata}
\label{sec-verif} 

In this section we settle the complexity of non-interference
verification problems for timed automata.

\subsection{SNNI verification}

The SNNI verification problem (SNNI-VP), asks to check whether a
system $\M$ is SNNI.

For timed automata, this problem has been proved to be
\emph{undecidable} in~\cite{gardey-secco-05} and the proof is based on
the fact that language containment for TA is
undecidable~\cite{AlurDill94}.  However, if we consider the subclass
of timed automata $\M$ such that $\M \backslash \Sigma_h$ is
\emph{deterministic}, then the problem becomes decidable.  In the
sequel, we called $\TAldet$ the class of timed automata $\M$ such that
$\M \backslash \Sigma_h$ is deterministic.

\begin{theorem}\label{thm-snnivp}
  The SNNI-VP is PSPACE-complete for $\TAldet$. 
\end{theorem}

\begin{IEEEproof}
  Let $\M_1$ and $\M_2$ be two timed automata.  Checking whether
  $\lang(\M_2) \subseteq \lang(\M_1)$ with $\M_1$ a deterministic TA
  is PSPACE-complete~\cite{AlurDill94}.
%$\lang(\M_1 \backslash
%  \Sigma_h) \subseteq \lang(\M_1 / \Sigma_h)$ is always true. 
  Checking $\lang(\M / \Sigma_h) \subseteq \lang(\M \backslash
  \Sigma_h)$ can thus be done is PSPACE if $\M \backslash \Sigma_h$ is
  deterministic. Using Proposition~\ref{prop-snni}, it follows that
  SNNI-VP is PSPACE-easy for $\TAldet$.

%  \FC{I think we should
%     add that $h$ resets all the clocks in $A_2$.} 
%     \OR{Non, par contre il faut obliger h a se produire a la date 0 avec une nouvelle horloge $x_h$ et une garde $x_h=0$. Je change ca. }
% \FC{FC: oui je suis OK mais il faut aussi ne pas pouvoir repasser par $q_{01}$ sinon on melange
% les deux langages; j'ai ajout2 cela.}
%     \OR{Je pense que ca ne sert \`a rien car j'avais mis la garde $x_h=0$ pour $h$  (et $x_h$ n'est jamais reset\`ee) donc si on fait un mot $w$ dans $A_1$, que l'on revient \`a l'etat initial et que $h$ est encore franchissable, c'est que ce mot $w$ est en temps nul et que si il y a un mot $w'$ dans le langage il y a aussi le mot $w^*.w'$}
\begin{figure}
\begin{center}
  \begin{tikzpicture}[thick,node distance=5cm,every state/.style={draw=none}]%
    \node[state,initial] (q012)  {$q^0_{12}$}; 
    \node[below of=q012,yshift=4.5cm] (inv) {$[x \leq 0]$};
    \node[state,right of=q012] (q0)  {$q_{01}$}; 
    \node[state] (q1) [above right of=q0] {$q_{02}$};
    \path[->] (q012)  edge  node  {$h$} (q1);
    \path[->] (q012)  edge  node  {$\varepsilon$} (q0);
    \draw[blue] (q0) circle (1.5cm); 
% \node[draw=blue,inner sep=1pt,thick,ellipse,xshift=-.2cm,fit=(q0) (q2)] {}; 
 \node (x) [right of =q0,shift={(-3cm,-.2cm)}] {\textcolor{blue}{$\M_1$}};
 %\node[draw=red,inner sep=3pt,thick,ellipse,fit=(q1) (q3)] {}; 

 \draw[red] (q1) circle (1.5cm); 
 \node (y) [right of =q1,shift={(-3cm,-.1cm)}] {\textcolor{red}{$\M_2$}};
\end{tikzpicture}
\caption{The timed automaton $\M_{12}$}
\label{fig-snnivpproof}
\end{center}
\end{figure}
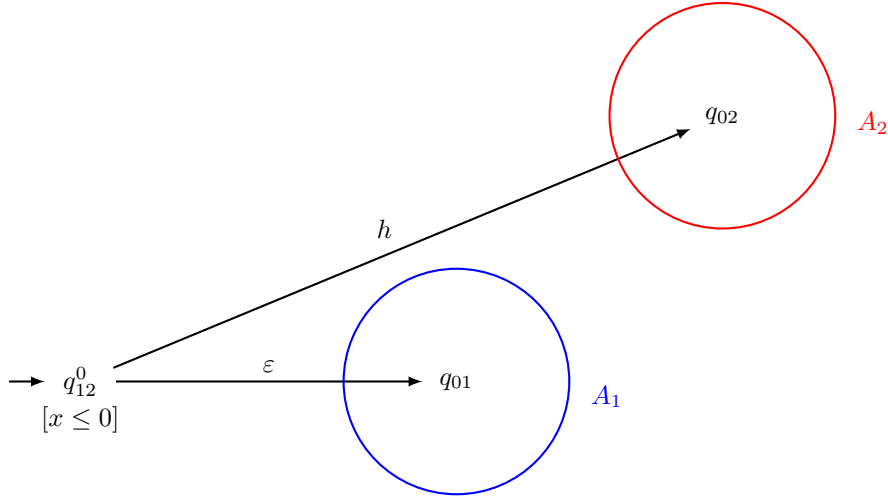

For PSPACE-hardness, we reduce the language inclusion problem
$\lang(\M_2) \subseteq \lang(\M_1)$, with $\M_1$ a deterministic TA,
to the SNNI-VP.  Let $\M_1=(Q_1,q_{01},X_1,\Sigma,E_1,Inv_1)$ be a
deterministic TA and $\M_2=(Q_2,q_{02},X_2,\Sigma,E_2,Inv_2)$ a
TA\footnote{We assume that $Q_1 \cap Q_2 = \emptyset$ and $X_1 \cap
  X_2=\emptyset$.}. We let $h \not \in \Sigma$ be a fresh letter, $x
\not \in X_1 \cup X_2 $ be a fresh clock and define
$\M_{12}=(\{q^0_{12}\} \cup Q_1 \cup Q_2,q_{01},X_1 \cup X_2 \cup
\{x\},\Sigma^{\varepsilon} \cup \{h\},E_{12},Inv_{12})$ be the timed
automaton defined (as shown in figure~\ref{fig-snnivpproof}) as
follows:
\begin{itemize}
\item the transition relation $E_{12}$ contains $E_1 \cup E_2$ and the
  additional transitions $(q^0_{12},true,h,\emptyset,q_{02})$ and
  $(q^0_{12},true,\varepsilon,\emptyset,q_{01})$;
%$E_i \subseteq E_{12}$,  $\forall i \in \{1,2\}$,
\item $Inv_{12}(q)=Inv_i(q)$ if $q \in Q_i, i \in \{1,2\}$, and
  $Inv_{12}(q^0_{12})=[x\leq 0]$.
% \item $(q_{01},\true,h,\emptyset,q_{02}) \in E_{12}$ 
\end{itemize}
We let $\Sigma_l=\Sigma$ and $\Sigma_h=\{h\}$.
We prove that $\M_{12}$ is SNNI iff $\lang(\M_2) \subseteq \lang(\M_1)$. 
This is easily established as:
\begin{eqnarray*}
  A_{12} \text{ is SNNI}  & \text{iff} & 
  \lang(\M_{12}/ \Sigma_h) \subseteq \lang(\M_{12} \backslash \Sigma_h) \hspace*{2cm} \text{[Proposition~\ref{prop-snni}]}\hspace{1em} \\
  &  \text{iff} & \lang(A_1) \cup \lang(A_2) \subseteq \lang(A_1) \\
  &  \text{iff} & \lang(A_2) \subseteq \lang(A_1)\mathpunct.
\end{eqnarray*}
% First if $\M_{12}$ is SNNI then $\lang(\M_2) \subseteq \lang(\M_1)$.
% Indeed, $\lang(\M_{12} \backslash \Sigma_h)=\lang(A_1)$ and
% $\lang(\M_{12}/\Sigma_h)=\lang(A_1) \cup \lang(A_2)$.  Hence
% $\lang(A_2) \subseteq \lang(A_1)$.

%  because
% if $\M_{12}$ is SNNI, $\lang(\M_2)\subseteq \lang(\M_2) \cup
% \lang(\M_1)=\lang(\M_{12} /\{h\})=\lang(\M_{12} \backslash
% \{h\})=\lang(\M_1)$.  $\lang(\M_2) \subseteq \lang(\M_1) \ \Rightarrow
% \M_{12}$ is SNNI because if $\lang(\M_2)\subseteq \lang(\M_1)$ then
% $\lang(\M_{12} /\{h\})=\lang(\M_2) \cup \lang(\M_1) \\ \subseteq
% \lang(\M_1)=\lang(\M_{12} \backslash \{h\})$ and then $\lang(\M_{12}
% /\{h\})=\lang(\M_{12} \backslash\{h\})$.
Thus the SNNI-VP is
PSPACE-complete for $\TAldet$.
\end{IEEEproof}

For non-deterministic finite automata $A_1$ and $A_2$, checking
language inclusion $\lang(A_1) \subseteq \lang(A_2)$ is
PSPACE-complete~\cite{stockmeyer-73}. Then, using the same proof with
$\M_1$ being a non deterministic finite automaton, It follows that:
\begin{corollary}\label{snni-vp-fndta}
The SNNI-VP is PSPACE-complete for non-deterministic finite automata.
\end{corollary}
%\begin{corollary} \label{snni-vp-f}
% For finite automata, SNNI-VP is PSPACE-complete. 
% \end{corollary}
Moreover, when $A_2$ is a deterministic finite automaton, language containment
can be checked in PTIME and thus we have the following corollary:
\begin{corollary}\label{snni-vp-fdta}
  For finite automata belonging to $\TAldet$, the SNNI-VP is PTIME.
\end{corollary}

%  and in PTIME
% for deterministic. Table 1 summarizes the results for the SNNI-VP.

% For finite automata SNNI-VP si decidable since it can be reduced to a
% trace equivalence problem (see the proof of theorem~\ref{thm-snnivp}),
% and we can deduce the following corollaries:

%  \begin{corollary} \label{snni-vp-fdta}
%  For finite automata belonging to $\TAldet$, the  SNNI-VP is PTIME.
%  \end{corollary}
 
% \subsubsection{Summary of the results}

The table~\ref{tab-res-snni-vp} summarizes the results on the
complexity of the SNNI-VP.

\begin{table}[thbtp]
  \centering
  \begin{tabular}{||l||c|c||}
    \cline{1-3}%\cline{2-3}
    \multicolumn{1}{||c||}{} & \multicolumn{1}{c|} {Timed Automata} & {~~ Finite Automata~}  \\ \hline\hline
    $\M \backslash \Sigma_h$ is \emph{deterministic}~($\TAldet$)   & ~~PSPACE-complete~(Theorem~\ref{thm-snnivp}) & PTIME ~(Corollary~\ref{snni-vp-fdta}) \\\hline 
    ~General Case ~~  & Undecidable~\cite{gardey-secco-05} & PSPACE-complete (Corollary~\ref{snni-vp-fndta})  \\\hline\hline
  \end{tabular}
  \smallskip
  \caption{Complexity if SNNI-VP}
  \label{tab-res-snni-vp}
\end{table}

\subsection{Verification of CSNNI and BSNNI properties}

BSNNI-VP and CSNNI-VP are decidable for timed
automata~\cite{gardey-secco-05} since simulation and bisimulation are
decidable. % for timed automata.
For finite automata, the complexity of BSNNI-VP and CSNNI-VP is known
to be PTIME~\cite{cassez-mmm-07}. We settle here the complexity of
those problems for timed automata.

\begin{theorem}\label{thm-bcsnnivp}
  The CSNNI-VP and BSNNI-VP are EXPTIME-complete for Timed Automata.
\end{theorem}
\begin{IEEEproof}
Strong timed bisimilarity and simulation pre-order are both
  EXPTIME-complete for timed automata.  The EXPTIME-hardness is
  established in~\cite{laroussinie-FOSSAC-00} where it is shown that
  any relation between simulation pre-order and bisimilarity is
  EXPTIME-hard for Timed Automata.

  The EXPTIME-easiness for strong timed bisimulation was established
  in~\cite{Cerans92} and for simulation pre-order
  in~\cite{tasiran-concur-96}.

  To establish EXPTIME-completeness for CSNNI-VP and BSNNI-VP, we show
  that these problems are equivalent to their counterparts for timed
  automata.

  To do this, we use the automata $A_1,A_2$ and $A_{12}$ already
  defined in the proof of Theorem~\ref{thm-snnivp}.
  
  We show that: $A_1$ simulates $A_2$ iff $A_{12}$ is CSNNI.

  Assume $A_1$ simulates $A_2$. There exists a relation $\R$ \st: 1)
  $(q_{01},\vec{0}_{X_1}) \R (q_{01},\vec{0}_{X_1})$ and 2) for each
  state $(s_2,\vec{x_2})$, there exists $ (s_1,\vec{x_1})$ \st
  $(s_2,\vec{x_2}) \R (s_1,\vec{x_1})$, and whenever $(s_2,\vec{x_2})
  \xrightarrow{\ a\ } (s'_2,\vec{x_2}')$ for $a \in \Sigma \cup
  \setRp$, then $(s_1,\vec{x_1}) \xrightarrow{\ a\ } (s'_1,\vec{x_1}')$
  and $(s'_2,\vec{x_2}') \R (s'_1,\vec{x_1}')$.

  We define a relation $\R'$ for each $(\ell,\vec{x_1}\vec{x_2}x)$ of
  $A_{12} / \Sigma_h$ to a state $(\ell',\vec{x_1}'\vec{x_2}'x')$ of
  $A_{12} \backslash \Sigma_h$ as follows:
%let $(\ell,\vec{x_1}\vec{x_2}x)$ be a state of $A_{12} /
%  \Sigma_h$, then
  \begin{itemize}
  \item if $\ell=q^0_{12}$ then $(\ell,\vec{x_1}\vec{x_2}x) \R'
    (\ell,\vec{x_1}'\vec{x_2}'x')$;
  \item if $\ell \in Q_1$, then $(\ell,\vec{x_1}\vec{x_2}x) \R'
    (\ell,\vec{x_1}\vec{x_2}'x')$;
  \item if $\ell \in Q_2$,  then $(\ell,\vec{x_1}\vec{x_2}x) \R'
    (\ell',\vec{x_1}'\vec{x_2}'x')$ iff $(\ell,\vec{x_2}) \R
    (\ell',\vec{x_1})$;
  \end{itemize}
  $\R'$ is a simulation of $A_{12} / \Sigma_h$ by $A_{12} \backslash
  \Sigma_h$:
  \begin{itemize}
  \item the initial states of the two TA are in relation;
  \item assume $(s,\vec{x_1}\vec{x_2}x) \xrightarrow{\ a\ }_{A_{12} /
      \Sigma_h} (s',\vec{x_1}'\vec{x_2}'x')$; If $s \in \{q_{12}^0\}
    \cup Q_1$ then clearly it is simulated by the same state in
    $A_{12} \backslash \Sigma_h$ . Otherwise, if $s \in Q_2$, then
    there exists a state $(\ell',\vec{x_1}\vec{x_2}'x')$ in $A_{12}
    \backslash \Sigma_h$ \st $(s,\vec{x_1}\vec{x_2}x) \R'
    (s',\vec{x_1}'\vec{x_2}'x')$: by definition of $\R'$ we can take
    any $(s',\vec{x_1}'\vec{x_2}'x')$ with $(s,\vec{x_2}) \R
    (s',\vec{x_1}')$. It is easy to see that because $A_1$ can
    simulate $A_2$ from there on, $\R'$ is indeed a simulation
    relation. Thus $A_{12} / \Sigma_h$ and $A_{12} \backslash
    \Sigma_h$ are co-similar by Proposition~\ref{def-snni2}.
  \end{itemize}
 
  Now assume conversely that there is a simulation $\R'$ of $A_{12} /
  \Sigma_h$ by $A_{12} \backslash \Sigma_h$.  We can define a
  simulation relation of $A_2$ by $A_1$ as follows: each state
  $(s,\vec{x_1}\vec{x_2}x)$ with $s \in Q_2$ of $A_{12} / \Sigma_h$ is
  simulated by a state $(s',\vec{x_1}'\vec{x_2}'x')$ with $s' \in
  Q_1$.  We then define $\R$ by $(s,\vec{x_2}) \R (s',\vec{x_1}')$.
  Again it is easy to see that $\R$ is a simulation relation.

  It follows that CSNNI is EXPTIME-complete.

  \medskip

  Now assume that $A_1$ and $A_2$ are bisimilar. We can define the
  relation $\R'$ exactly as above and this time it is a weak
  bisimulation between $A_{12} \backslash \Sigma_h$ and
  $A_{12}/\Sigma_h$.

  If $A_{12}$ is BSNNI, the bisimulation relation $\R'$ between
  $A_{12} \backslash \Sigma_h$ and $A_{12}/\Sigma_h$ induces a
  bisimulation relation $\R$ between $A_1$ and $A_2$: it suffices to
  build $\R$ as the restriction of $\R'$ between states with a
  discrete component in $Q_1$ and a discrete component in $Q_2$.

  As checking bisimulation between TA is also EXPTIME-complete, the
  EXPTIME-completeness of BSNNI-VP for TA follows.

\end{IEEEproof}

The table~\ref{tab-res-bcsnni-vp} summarize the results on the verification of the CSNNI and BSNNI properties. 

\begin{table}[hbtp]
  \centering
  \begin{tabular}{||l||c|c||}
    \cline{1-3}%\cline{2-3}
    \multicolumn{1}{||c||}{} & \multicolumn{1}{c|} {Timed Automata} & {~~ Finite Automata~}  \\ \hline\hline
   CSNNI-VP  ~~  & ~~EXPTIME-C~(Theorem~\ref{thm-bcsnnivp}) & PTIME~\cite{cassez-mmm-07} \\\hline 
    BSNNI-VP ~~  & ~~EXPTIME-C~(Theorem~\ref{thm-bcsnnivp}) & PTIME~\cite{cassez-mmm-07} \\\hline\hline
  \end{tabular}
  \smallskip
  \caption{Results for CSNNI-VP and BSNNI-VP}
  \label{tab-res-bcsnni-vp}
\end{table}

\section{The SNNI Control Problem}
\label{sec-results}

The previous non-interference verification problem, consists in
\emph{checking} whether an automaton $A$ has the non-interference
property. If the answer is ``no'', one has to investigate why the
non-interference property is not true, modify $A$ and check the
property again. In contrast to the verification problem, the synthesis
problem indicates whether there is a way of restricting the behavior
of users to ensure a given property.  Thus we consider that only some
actions in the set $\Sigma_c $, with $\Sigma_c \subseteq \Sigma_h \cup
\Sigma_l$, are controllable and can be disabled.  We let $\Sigma_u =
\Sigma \setminus \Sigma_c$ denote the actions that are uncontrollable
and thus cannot be disabled.  Note that, contrary
to~\cite{cassez-mmm-07}, we release the constraint $\Sigma_c
=\Sigma_h$.  The motivations for this work are many fold.  Releasing
$\Sigma_c =\Sigma_h$ is interesting in practice because it enables one
to specify that an action from $\Sigma_h$ cannot be disabled (a
service must be given), while some actions of $\Sigma_l$ can be
disabled. We can view actions of $\Sigma_l$ as capabilities of the
low-level user (\eg pressing a button), and it thus makes sense to
prevent the user from using the button for instance by
disabling/hiding it temporarily.

Recall that a \emph{controller} $C$ for $A$ gives for each run $\rho$
of $A$ the set $C(\rho) \in 2^{\Sigma_c \cup \{\lambda\}}$ of actions
that are enabled after this particular run.
The SNNI-\emph{Control Problem} (SNNI-CP) we are interested in is the
following:
\begin{equation}\label{eq-ssni-rp} \tag{SNNI-CP}
  \textit{Is there a controller $C$  s.t. $C(A)$ is SNNI ?}
\end{equation}

The SNNI-\emph{Controller Synthesis Problem} (SNNI-CSP) asks to
compute a witness when the answer to the SNNI-CP is ``yes''.

\subsection{Preliminary Remarks}
\begin{figure}  \centering
\begin{tikzpicture}[thick,node distance=1.3cm and 2.3cm,initial text=,auto,scale=0.7]% 
    \node[state,initial] (0) {$0$}; 
    \node[state] (3) [below=of 0] {$3$};
    \node[state] (1) [right=of 0] {$1$};
    \node[state] (2) [below=of 1] {$2$};
    \path[->] (0)  edge [swap] node  {$a$} (3)
                    edge  node  {$h$} (1)
              (1) edge  node  {$a$} (2);
  \end{tikzpicture}
\caption{Automaton $D$}
\label{fig-most-subnotmost}
\end{figure}
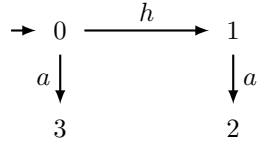
\noindent First we motivate our definition of controllers which are
mappings from $\runs(A)$ to $2^{\Sigma_c \cup \{\lambda\}}$.  The
common definition of a controller in the literature is a mapping from
$\runs(A)$ to $\Sigma_c \cup \{\lambda\}$.  Indeed, for the safety (or
reachability) control problem, one can compute a mapping $M : \runs(A)
\rightarrow 2^{\Sigma_c \cup \{\lambda\}}$ (most permissive
controller), and a controller $C$ ensures the safety goal iff $C(\rho)
\in M(\rho)$.  This implies that any sub-controller of $M$ is a good
controller.  This is not the case for SNNI, even for finite automata,
as the following example shows.

\medskip

\begin{example}
  Let us consider the automaton $D$ of
  Figure~\ref{fig-most-subnotmost} with $\Sigma_c=\{a,h\}$.  The
  largest sub-system of $D$ which is SNNI is $D$ itself.  Disabling
  $a$ from state $0$ will result in an automaton which is not
  SNNI.
\end{example}

We are thus interested in computing the largest (if there is such)
sub-system of $A$ that we can control which is SNNI.
Second, in our definition we allow a controller to forbid any
controllable action. In contrast, in the literature, a controller
should ensure some liveness and never block the system.  In the
context of security property, it makes sense to disable everything if
the security policy cannot be enforced otherwise. This makes the SNNI-CP easy for finite automata.

\subsection{SNNI-VP versus SNNI-CP}

SNNI-CP is harder than SNNI-VP since SNNI-VP reduces to SNNI-CP by taking $\Sigma_c=\emptyset$.
Note that this is not true if we restrict to the subclass of control where $\Sigma_c = \Sigma_h$. Indeed, in this case SNNI-CP is always true (and then decidable) since the controller which forbid all controllable transitions make the system SNNI. 

We then have the following  theorem:

\begin{theorem}\label{thm-undecidable}
For general Timed Automata, SNNI-CP and SNNI-CSP are undecidable.
\end{theorem}
\begin{IEEEproof}
SNNI-CP obviously reduces to SNNI-CSP. SNNI-VP reduces to SNNI-CP by taking $\Sigma_c=\emptyset$.
SNNI-VP is undecidable for non-deterministic Timed Automata.

\end{IEEEproof}

We will now show that SNNI-CP reduces to the SNNI-VP for finite automata.

\begin{theorem}\label{thm-0}
  For finite automata, the SNNI-CP is PSPACE-Complete.
\end{theorem}
\begin{IEEEproof}
The proof %(Appendix~\ref{app-proofs})
consists in proving that if a finite automaton can be restricted to be
SNNI, then disabling all the $\Sigma_c$ actions is a solution. Thus
the SNNI-CP reduces to the SNNI-VP and the result follows.

  As time is not taken into account in untimed automaton, we can have
  $C(\rho)=\emptyset$ for finite automaton (for general timed
  automaton, this would mean that we block the time.)  The proof of
  the theorem consists in proving that if a finite automaton can be
  restricted to be SNNI, then disabling all the $\Sigma_c$ actions is
  a solution. Let $C_\forall$ be the controller defined by
  $C_\forall(\rho)=\emptyset$. We prove the following: if $C$ is a
  controller \st $C(A)$ is SNNI, then $C_\forall(A)$ is SNNI.

  Assume a finite automaton $D$ is SNNI. Let $e \in \Sigma_h \cup
  \Sigma_l$ and let $\lang_e$ be the set of words containing at least
  one $e$. Depending on the type of $e$ we have:
\begin{itemize}
\item if $e \in \Sigma_l$, then $\lang((D \backslash \{e\}) \backslash
  \Sigma_h)=\lang(D \backslash \Sigma_h) \setminus \lang_e$ and as $D$
  is SNNI, it is also equal to $\lang(D / \Sigma_h) \setminus \lang_e
  = \lang((D \backslash \{e\}) / \Sigma_h)$;
\item if $e \in \Sigma_h$, $\lang((D \backslash \{e\})/\Sigma_h) \subseteq \lang(D /
  \Sigma_h) =\lang(D \backslash \Sigma_h)= \lang((D \backslash \{e\}) \backslash
  \Sigma_h)$.
\end{itemize}
So, if $D$ is SNNI, $D \backslash L$ is SNNI, $\forall L \subseteq
\Sigma$. Since $\lang(C_\forall(D)) =
\lang(D \backslash \Sigma_c)$, if $D$ is SNNI, then $D \backslash
\Sigma_c$ is also SNNI and therefore $C_\forall(D)$ is SNNI.

Let $A$ be the TA we want to restrict. Assume there is a controller
$C$ \st $C(A)$ is SNNI. $C_\forall(C(A))$ is SNNI so
$C_\forall(C(A))=C_\forall(A)$ is also SNNI which means that $A
\backslash \Sigma_c$ is SNNI. This proves that: $\exists C$ \st $C(A)$
is SNNI $\Leftrightarrow$ $A \backslash \Sigma_c$ is SNNI.

It is then equivalent to check that $A \backslash \Sigma_c$ is SNNI to
solve the SNNI-CP for $A$ and this can be done in
PSPACE. PSPACE-hardness comes from the reduction of SNNI-VP to
SNNI-CP, by taking $\Sigma_c=\emptyset$.  

\end{IEEEproof}

Moreover since the SNNI-CP reduces to the SNNI-VP for finite automata, and from corollary~\ref{snni-vp-fdta} we have the following result:
\begin{corollary}\label{cor-SNNI-CP-fdta}
  For finite automata belonging to $\TAldet$, the SNNI-CP is PTIME.
\end{corollary}

We will now show that Theorem~\ref{thm-0} does not hold for timed automata as the following
example demonstrates.

\begin{example} \label{ex-snni2} 
  Figure~\ref{fig-snni-ex2} gives an example of a timed automaton $H$
  with high-level actions $\Sigma_h=\{h\}$ and low-level actions
  $\Sigma_l=\{a,b\}$.

\begin{figure}  \centering
\begin{tikzpicture}[thick,node distance=1.3cm and 2.4cm,initial text=,auto,scale=0.7]% 
%\begin{tikzpicture}[thick,node distance=1.3cm and 2.4cm]% 
    \node[state,initial] (a) {$0$}; 
    \node[state] (b) [below=of a] {$1$};
    \node[state] (c) [right=of a] {$2$};
    \node[state] (d) [below=of c] {$3$};
    \path[->] (a)  edge  [swap] node  {$a,x > 1$} (b)
                    edge  [dashed] node  {$h,x >4$} (c)
              (c) edge   [dashed] node  {$b$} (d);
  \end{tikzpicture}
\caption{The Automaton $H$}
\label{fig-snni-ex2}
\end{figure}
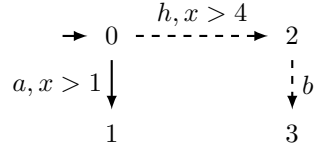
\noindent Assume $\Sigma_c=\{a\}$. Notice that $H \backslash \Sigma_c$
is not SNNI. Let the state based controller $C$ be de\-fi\-ned by:
$C(0,x)=\{a,\lambda\}$ when $H$ is in state $(0,x)$ with $x < 4$; and
$C(0,x)=\{a\}$ when $x=4$.
Then $C(H)$ is SNNI.  In this example, when $x=4$ we prevent time from
elapsing by forcing the firing of $a$ which indirectly disables action
$h$. To do this we just have to add an invariant $[x \leq 4]$ to
location $0$ of $H$ and this cuts out the dashed transitions rendering
$C(H)$ SNNI.
\end{example}

\subsection{Algorithms for SNNI-CP and SNNI-CSP}
In this section we first prove that the SNNI-CP is EXPTIME-hard for
$\TAldet$.  Then we give an EXPTIME algorithm to solve the SNNI-CP and
SNNI-CSP.

\begin{theorem}\label{thm-00}
  For $\TAldet$, the SNNI-CP is EXPTIME-Hard.
\end{theorem}
\begin{IEEEproof}
  The safety control problem for TA is
  EXPTIME-hard~\cite{HenzingerKopke97}.  In the proof of this theorem,
  T.A.~Henzinger and P.W.~Kopke use timed automata where the
  controller chooses an action and the environment resolves
  non-determinism. The hardness proof reduces the halting problem for
  alternating Turing Machines using polynomial space to a safety
  control problem.  In our framework, we use TA with controllable and
  uncontrollable actions.  It is not difficult to adapt the hardness
  proof of~\cite{HenzingerKopke97} to TA which are deterministic \wrt
  $\Sigma_c$ actions and non deterministic \wrt $\Sigma_u$ actions.
  As $\Sigma_u$ transitions can never be disabled (they act only as
  spoiling actions), we can use a different label for each
  uncontrollable transition without altering the result in our
  definition of the safety control problem. Hence: the safety control
  problem as defined in section~\ref{sec-prelim} is EXPTIME-hard for
  deterministic TA (with controllable and uncontrollable
  transitions). This problem can be reduced to the safety control
  problem of TA with only one state $bad$.
  We can now reduce the safety control problem for deterministic TA
  which is EXPTIME-hard to the SNNI control problem on $\TAldet$. Let
  $A=(Q \cup \{bad\},q_0,X,$ $\Sigma_c \cup \Sigma_u,E,Inv)$ be a TGA,
  with $\Sigma_c$ (resp. $\Sigma_u$) the set of controllable
  (resp. uncontrollable) actions, and $bad$ a location to avoid.  We
  define $A'$ by adding to $A$ two uncontrollable transitions:
  $(bad,\true,h,\emptyset,q_h)$ and $(q_h,\true,l,\emptyset,q_l)$
  where $q_h$ and $q_l$ are fresh locations with invariant
  $\true$. $l$ and $h$ are two fresh uncontrollable actions in
  $A'$. We now define $\Sigma_h=\{h\}$ and $\Sigma_l=\Sigma_c \cup
  \Sigma_u \cup \{l\}$ for $A'$.  By definition of $A'$, for any
  controller $C$, if location $\bad$ is not reachable in $C(A')$, then
  the actions $h$ and then $l$ can not be fired. Thus if there is
  controller for $C$ for $A$ which avoids $\bad$, the same controller
  $C$ renders $A'$ SNNI.  Now if there is a controller $C'$ \st
  $C'(A')$ is SNNI, it must never enable $h$: otherwise a (untimed)
  word $w.h.l$ would be in $\untimed(\lang(C'(A') / \Sigma_h))$ but as
  no untimed word containing an $l$ can be in $\untimed(\lang(C'(A')
  \backslash \Sigma_h))$, and thus $C'(A')$ would not be
  SNNI. 
  Notice that it does not matter whether we require the controllers to
  be non blocking (mappings from $\runs(A)$ to $2^{\Sigma_c \cup
    \{\lambda\}} \setminus \emptyset$) or not as the reduction holds
  in any case.  
\end{IEEEproof}

%\vspace{0em plus 1em}

To compute the most permissive controller (and we will also prove
there is one), we build a safety game and solve a safety control
problem. It may be necessary to iterate this procedure.  Of course, we
restrict our attention to TA in the class $\TAldet$ for which the
SNNI-VP is decidable.

%bb
Let $A=(Q,q_0,X,\Sigma_h \cup \Sigma_l,E,Inv)$ be a TA \st $A
\backslash \Sigma_h$ is deterministic.  The idea of the reduction
follows from the following remark: we want to find a controller $C$
\st $\lang(C(A) \backslash \Sigma_h)=\lang(C(A) / \Sigma_h)$.  For any
controller $C$ we have $\lang(C(A) \backslash \Sigma_h) \subseteq
\lang(C(A) / \Sigma_h)$ because each run of $C(A) \backslash
\Sigma_h$ is a run of $C(A)/ \Sigma_h)$.  To ensure SNNI we must have
$\lang(C(A) / \Sigma_h) \subseteq \lang(A \backslash \Sigma_h)$:
indeed, $A \backslash \Sigma_h$ is the largest language that can be
generated with no $\Sigma_h$ actions, so a necessary condition for
enforcing SNNI is $\lang(C(A) / \Sigma_h) \subseteq \lang(A \backslash
\Sigma_h)$. The controller $C(A)$ indicates what must be pruned out in
$A$ to ensure the previous inclusion.  Our algorithm thus proceeds as
follows: we first try to find a controller $C^1$ which ensures that
$\lang(C^1(A) / \Sigma_h) \subseteq \lang(A \backslash \Sigma_h)$. If
$\lang(C^1(A) / \Sigma_h) = \lang(A \backslash \Sigma_h)$ then $C^1$
is the most permissive controller that enforces SNNI. It could be that
what we had to prune out to ensure $\lang(C^1(A) / \Sigma_h) \subseteq
\lang(A \backslash \Sigma_h)$ does not render $C^1(A)$ SNNI. In this
case we may have to iterate the previous procedure on the new system
$C^1(A)$.

We first show how to compute $C^1$. As $A \backslash \Sigma_h$ is
deterministic, we can construct $A_2=(Q \cup
\{q_{bad}\},q_0^2,X_2,\Sigma_h \cup \Sigma_l,E_2,Inv_2)$ which is a
copy of $A$ (with clock renaming) with $q_{bad}$ being a fresh
location and \st $A_2$ is a \emph{complete} (\ie
$\lang(A_2)=T\Sigma^*$) version of $A \backslash \Sigma_h$ ($A_2$ is
also deterministic).  We write $\last_2(w)$ the state $(q,v)$ reached
in $A_2$ after reading a timed word $w \in T\Sigma^*$.  $A_2$ has the
property that $w \in \lang(A \backslash \Sigma_h)$ if the state
reached in $A_2$ after reading $w$ is not in $\bad$ with
$\bad=\{(q_{bad},v) \ | \ v \in {\setRp^X}\}$.
\begin{fact}\label{fact-1}
  Let $w \in T\Sigma^*$. Then $w \not\in \lang(A \backslash
  \Sigma_h) \iff \last_2(w) \in \bad$.
\end{fact}
\noindent We now define the product $A_p=A \times_{\Sigma_l} A_2$ and
the set of bad states, $\bad^\otimes$ of $A_p$ to be the set of states
where $A_2$ is in $\bad$.  $\xrightarrow{}_p$ denotes the transition
relation of the semantics of $A_p$ and $s_p^0$ the initial state of
$A_p$. When it is clear from the context we omit the subscript $p$ in
$\xrightarrow{}_p$.
\noindent
\begin{lemma}\label{lemma-2}
  Let $w \in \lang(A)$.  Then there is a run $\rho \in \runs(A_p)$
  \st $\rho=s_p^0 \xrightarrow{\ w\ }_p s$ with $s \in \bad^\otimes$ iff
  $\proj{\Sigma_l}(w) \not\in \lang(A \backslash \Sigma_h)$.
\end{lemma}
The proof follows easily from Fact~\ref{fact-1}.
Given a run $\rho$ in $\runs(A_p)$, we let $\rho_{|1}$ be the
projection of the run $\rho$ on $A$ (uniquely determined) and
$\rho_{|2}$ be the unique run\footnote{Recall that $A_2$ is
  deterministic.} in $A_2$ whose trace is
$\proj{\Sigma_l}(\trace(\rho))$.  The following Theorem proves that
any controller $C$ \st $C(A)$ is SNNI can be used to ensure that
$\bad^\otimes$ is not reachable in the game $A_p$:
\begin{lemma}\label{lem-thm-1}
  Let $C$ be a controller for $A$ \st $C(A)$ is SNNI.  Let $C^\otimes$
  be a controller on $A_p$ defined by
  $C^\otimes(\rho')=C(\rho'_{|1})$.  Then, $\reach(C^\otimes(A_p))
  \cap \bad^\otimes = \emptyset$.
\end{lemma}
\begin{IEEEproof}
  First $C^\otimes$ is well-defined because $\rho'_{|1}$ is uniquely
  defined.  Let $C$ be a controller for $A$ \st $C(A)$ is SNNI.
  Assume $\reach(C^\otimes(A_p)) \cap \bad^\otimes \neq
  \emptyset$. By definition, there is a run $\rho'$ in
  $\runs(C^\otimes(A_p))$ such that:
\begin{eqnarray*}
\rho' & = & ((q_0,q_0^2),(\vec{0},\vec{0})) \xrightarrow{\, e_1 \,}
  ((q_1,q'_1),(v_1,v'_1)) \xrightarrow{\, e_2 \,} \cdots
  \xrightarrow{\, e_n \,} ((q_n,q'_n),(v_n,v'_n)) \\ & &  \xrightarrow{\,
    e_{n+1} \,} ((q_{n+1},q'_{n+1}),(v_{n+1},v'_{n+1})) 
\end{eqnarray*}
with $((q_{n+1},q'_{n+1}),(v_{n+1},v'_{n+1})) \in \bad^\otimes$ and we
can assume $(q'_i,v_i') \not \in \bad$ for $1 \leq i \leq n$ (and
$q_0^2 \not\in \bad$).  Let $\rho=\rho'_{|1}$ and $w=
\proj{\Sigma_l}(\trace(\rho')) = \proj{\Sigma_l}(\trace(\rho))$. We
can prove (1): $\rho \in \runs(C(A))$ and (2): $w \not\in
\lang(C(A)\backslash \Sigma_h)$.  (1) directly follows from the
definition of $C^\otimes$. This implies that $w \in
\lang(C(A)/\Sigma_h)$.  (2) follows from Lemma~\ref{lemma-2}. By (1)
and (2) we obtain that $w \in \lang(C(A)/\Sigma_h) \setminus
\lang(C(A)\backslash \Sigma_h)$ \ie $\lang(C(A)/\Sigma_h) \neq
\lang(C(A)\backslash \Sigma_h)$ and so $C(A)$ does not have the SNNI
property which is a contradiction. Hence $\reach(C^\otimes(A_p)) \cap
\bad^\otimes = \emptyset$. 
\end{IEEEproof}
If we have a controller which solves the safety game
$(A_p,\bad^\otimes)$, we can build a controller which ensures that
$\lang(C(A)/\Sigma_h) \subseteq \lang(A \backslash \Sigma_h)$.  Notice
that as emphasized before, this does not necessarily ensure that
$C(A)$ is SNNI.
\begin{lemma}\label{lem-thm-2}
  Let $C^\otimes$ be a controller for $A_p$ \st
  $\reach(C^\otimes(A_p)) \cap \bad^\otimes = \emptyset$. Let
  $C(\rho)=C^\otimes(\rho')$ if $\rho'_{|1}=\rho$. $C$ is well-defined
  and $\lang(C(A)/\Sigma_h) \subseteq \lang(A \backslash
  \Sigma_h)$.
\end{lemma}
\begin{IEEEproof}
  Let $\rho=(q_0,\vec{0}) \xrightarrow{\, e_1 \,} (q_1,v_1)
  \xrightarrow{\, e_2 \,} \cdots \xrightarrow{\, e_n \,} (q_n,v_n)$ be
  a run of $A$.  Since $A_2$ is deterministic and complete there is
  exactly one run $\rho'=((q_0,q_0),(\vec{0},\vec{0})) \xrightarrow{\,
    e_1 \,} ((q_1,q'_1),(v_1,v'_1)) \xrightarrow{\, e_2 \,} \cdots
  \xrightarrow{\, e_n \,} ((q_n,q'_n),(v_n,v'_n))$ in $A_p$ \st
  $\rho'_{|1}=\rho$.  So $C$ is well-defined.
  Now, assume there is some $w \in \lang(C(A)/\Sigma_h) \setminus
  \lang(A \backslash \Sigma_h)$.  Then, there is a run $\rho$ in
  $\runs(C(A)) \subseteq \runs(A)$ \st
  $\proj{\Sigma_l}(\trace(\rho))=w$, there is a unique run $\rho \in
  \runs(A_p)$ \st $\rho'_{|1}=\rho$ and $\trace(\rho')=w$.
  First by Lemma~\ref{lemma-2}, $\last(\rho') \in \bad^\otimes$.
  Second, this run $\rho'$ is in $\runs(C^\otimes(A_p))$ because of
  the definition of $C$.  Hence $\reach(C^\otimes(A_p)) \cap
  \bad^\otimes \neq \emptyset$ which is a contradiction.  
\end{IEEEproof}
It follows that if $C^\otimes$ is the most permissive controller for
$A_p$ then $C(A)$ is a timed automaton (and can be effectively
computed) because the most permissive controller for safety timed
games is memoryless.  More precisely, let $RG(A_p)$ be the the region
graph of $A_p$.  $C$ is memoryless on $RG(A_p \backslash \Sigma_h)$
because $A_2$ is deterministic. The memory required by $C$ is at most
$RG(A \backslash \Sigma_h)$ on the rest of the region graph of
$RG(A_p)$.

Assume the safety game $(A_p,\bad^\otimes)$ can be won and $C^\otimes$
is the most permissive controller.  Let $C$ be the controller obtained
using Lemma~\ref{lem-thm-2}.  Controller $C$ ensures that
$\lang(C(A)/\Sigma_h) \subseteq \lang(A \backslash \Sigma_h)$.  But as
the following example shows, it may be the case that $C(A)$ is not
SNNI.
\begin{example}
  \label{ex-counter-snni2} Consider the TA $K$ of
  Figure~\ref{fig-counter-snni} with $\Sigma_h=\{h\}$ and
  $\Sigma_c=\{a\}$.  

\begin{figure}  \centering
\begin{tikzpicture}[thick,node distance=1.3cm and 1.9cm,initial text=,auto,scale=0.7]% 
    \node[state,initial] (0) {$0$}; 
    \node[state] (1) [below=of 0] {$1$};
    \node[state] (2) [right=of 1] {$2$};
    \node[state] (3) [right=of 2] {$3$};
    \node[state] (4) [right=of 0] {$4$};
    \node[state] (5) [right=of 4] {$3$};
    \path[->] (0)  edge [dashed] node  {$a,x \geq 2$} (1)
                    edge  node  {$h$} (4)
              (1) edge [dashed] node  {$h$} (2)
              (2) edge [dashed] node  {$b$} (3)
              (4) edge  node  {$a, x \geq 2$} (5);
  \end{tikzpicture}
\caption{The Automaton $K$}
\label{fig-counter-snni}
\end{figure}
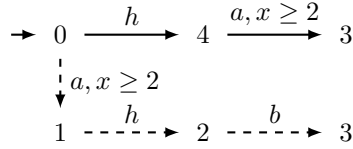
\noindent We can compute $C(K)$ from $C^\otimes$ which satisfies
$\reach(C^\otimes(K\times_{\Sigma_l} K_2))\cap \bad^\otimes =
\emptyset$, and is given by the sub-automaton of $K$ with the plain
arrows.  $C(K)$ is obviously not SNNI.  For the example of $A(1)$ in
Figure~\ref{fig-snni-ex1}, if we compute $C$ in the same manner, we
obtain $C(A(1))=A(2)$ and moreover $\lang(C(A(1))/\Sigma_h) =
\lang(A(1) \backslash \Sigma_h)$.  And then the most permissive
sub-system which is SNNI is given by $C(A(1))=A(2)$ (the guard $x \geq
1$ of $A(1)$ is strengthened). 
\end{example}

The example of Figure~\ref{fig-counter-snni} shows that computing the
most permissive controller on $A_p$ is not always sufficient.
Actually, we may have to iterate the computation of the most
permissive controller on the reduced system $C(A)$. 

\begin{lemma}\label{lemma-snniifequal}
  Consider the controller $C$ as defined in Lemma~\ref{lem-thm-2}. If
  $C(A)\backslash \Sigma_h \eqlang A \backslash \Sigma_h$ then $C(A)$
  is SNNI.
\end{lemma}
\begin{IEEEproof}
  If $C(A)\backslash \Sigma_h \eqlang A \backslash \Sigma_h$, then,
  $\lang(C(A)/\Sigma_h) \subseteq \lang(A \backslash
  \Sigma_h)=\lang(C(A) \backslash \Sigma_h)$. As $\lang(C(A)
  \backslash \Sigma_h) \subseteq \lang(C(A) / \Sigma_h)$ is always
  true, $\lang(C(A)/\Sigma_h) = \lang(C(A) \backslash \Sigma_h)$ and
  so, $C(A)$ is SNNI.  
\end{IEEEproof}

Let $\bot$ be the symbol that denotes non controllability (or the non
existence of a controller).
We inductively define the sequence of controllers $C^i$ and timed
automata $A^i$ as follows:
\begin{itemize}
\item let $C^0$ be the controller defined by $C^0(\rho)=2^{\Sigma_c
    \cup \{\lambda\}}$ and $A^0=C^0(A)=A$;
\item Let $A_{p}^i=A^i \times_{\Sigma_l} A^i_2$ and $C^\otimes_{i+1}$
  be the most permissive controller for the safety game
  $(A_{p}^i,\bad^\otimes_i)$ ($\bot$ if no such controller exists).
  We use the notation $\bad^\otimes_i$ because this set depends on
  $A^i_2$.  We define $C^{i+1}$ using Lemma~\ref{lem-thm-2}:
  $C^{i+1}(\rho)=C^\otimes_{i+1}(\rho')$ if $\rho'_{|1}=\rho$.  Let
  $A^{i+1}=C^{i+1}(A^i)$.
\end{itemize}
By Lemma~\ref{lemma-snniifequal}, if $C^{i+1}(A^{i})\backslash
\Sigma_h \eqlang A^i \backslash \Sigma_h$ then $C^{i+1}(A^i)$ is
SNNI. Therefore this condition is a sufficient condition for the
termination of the algorithm defined above:

\begin{lemma}\label{lem-thm-termination}
  There exists an index $i \geq 1$ \st $C^i(A^{i-1})$ is SNNI or
  $C^i=\bot$.
\end{lemma}
\begin{IEEEproof}
  We prove that the region graph of $C^{i+1}(A^i)$ is a sub-graph of
  the region graph of $C^1(A^0)$ for $i \geq 1$.  By
  Lemma~\ref{lem-thm-2} (and the remark following it), $C^1(A^0)$ is a
  sub-graph of $RG(A \times A_2)$. Moreover $C^1$ is memoryless on $A
  \backslash \Sigma_h$ and requires a memory of less than $|RG(A
  \backslash \Sigma_h)|$ on the remaining part. Assume on this part, a
  node of $RG(A \times A_2)$ is of the form $((q,r),k)$ where $q$ is a
  location of $A$ and $r$ a region of $A$ and $k \in \{1,|RG(A
  \backslash \Sigma_h)|\}$.

  Assume $RG(A^k)$ is a sub-graph of $RG(A^{k-1})$ for $k \geq 2$
  and $RG(A^{k-1} \backslash \Sigma_h)$ is sub-graph of $RG(A
  \backslash \Sigma_h)$. Using Lemma~\ref{lem-thm-2}, we can compute
  $A^k = C^{k}(A^{k-1})$ and: (1) $RG(A^k \backslash \Sigma_h)$ is a
  sub-graph of $A^{k-1} \backslash \Sigma_h$ and (2) the memory
  needed for $C^\otimes_k$ on the remaining part is less than
  $|RG(A^{k-1})|$. Actually, because $A^{k-1} \backslash \Sigma_h$ is
  deterministic, no more memory is required for $C^k$.  Indeed, the
  memory corresponds to the nodes of $A^{k} \backslash \Sigma_h$. Thus
  a node of $RG(A^k)$ which is not in $RG(A^{k} \backslash \Sigma_h)$
  is of the form $((q,r),k,k')$ with $k=k'$ or $k'=q_{bad}$. This
  implies that $RG(A^{k})$ is a sub-graph of $RG(A^{k-1})$.

  The most permissive controller $C^\otimes_i$ will either disable at
  least one controllable transition of $A_{p}^{i-1}\backslash
  \Sigma_h$ or keep all the controllable transitions of
  $A_{p}^{i-1}\backslash \Sigma_h$. In the latter case
  $A^{i}\backslash \Sigma_h=A^{i-1}\backslash \Sigma_h$ and otherwise
  $|RG(A^i\backslash \Sigma_h)| < |RG(A^{i-1}\backslash
  \Sigma_h)|$. This can go on at most $|RG(A\backslash \Sigma_h)|$
  steps. In the end either $A^{i}\backslash \Sigma_h=A^{i-1}\backslash
  \Sigma_h$ and this implies that $A^{i}\backslash \Sigma_h\eqlang
  A^{i-1}\backslash \Sigma_h$ (Lemma~\ref{lemma-snniifequal}) or it is
  impossible to control $A^{i-1}$ and $C^i=\bot$.
  In any case, our algorithm terminates in less than $|RG(A)|$ steps.  
\end{IEEEproof}

To prove that we obtain the most permissive controller which enforces
SNNI, we use the following Lemma:
\begin{lemma}\label{lemma-inclusion}
  If $M$ is a controller such that $\lang(M(A)/\Sigma_h) =
  \lang(M(A)\backslash \Sigma_h)$, then $ \forall i \geq 0$ and $
  \forall \rho \in \runs(A)$, $M(\rho) \subseteq C^{i}(\rho)$.
\end{lemma}

\begin{IEEEproof}The proof is by induction:
\begin{itemize}
\item for $i=0$ it holds trivially.
\item Assume the Lemma holds for indices up until $i$.  Thus we have
  $\runs(M(A)) \subseteq \runs(A^{i})$. Therefore, we can define $M$
  over $A^i$ and $M(A^i)$ is SNNI. By Lemma~\ref{lem-thm-1}, $M^\otimes$
  is a controller for the safety game $(A_{p}^i,\bad^\otimes_i)$,
  therefore $M^\otimes(\rho') \subseteq C^\otimes_{i+1}(\rho')$
  because $C^\otimes_{i+1}$ is the most permissive controller. This
  implies that $M(\rho) \subseteq C^{i+1}(\rho)$ by definition of
  $C^{i+1}$.  
\end{itemize}
\end{IEEEproof}
Using Lemma~\ref{lem-thm-termination}, the sequence $C^{i}$ converges to
a fix-point. Let $C^*$ denote this fix-point.
\begin{lemma}
  $C^*$ is the most permissive controller for the SNNI-CSP.
\end{lemma}
\begin{IEEEproof}
  Either $C^*=\bot$ and there is no way of enforcing SNNI
  (Lemma~\ref{lem-thm-1}), or $C^*\neq \bot$ is such that
  $\lang(C^*(A)/\Sigma_h) = \lang(C^*(A)\backslash \Sigma_h)$ by
  Lemma~\ref{lem-thm-2}. As for any valid controller $M$ such that
  $\lang(M(A)/\Sigma_h) = \lang(M(A)\backslash \Sigma_h)$ we have
  $M(\rho) \subseteq C^*(\rho)$ for each $\rho \in \runs(A)$
  (Lemma~\ref{lemma-inclusion}) the result follows.  
\end{IEEEproof}

% An example of how the previous algorithm works is given in
% Appendix~\ref{app-example-algo}.
%
Lemma~\ref{lem-thm-termination} proves the existence of a bound on
the number of times we have to solve safety games.
For a timed automaton $A$ in $\TAldet$, let $|A|$ be the size of $A$.
\begin{lemma}\label{lem-thm-bound-ta}
  For a $\TAldet$ $A$, $C^*$ can be computed in $O(2^{4.|A|})$.
\end{lemma}

\begin{IEEEproof}
  As the proof of Lemma~\ref{lem-thm-termination} shows, the region
  graph of $A^i$ is a sub-graph of the region graph of $A^1$, $\forall
  i \geq 1$, and the algorithm ends in less than $|RG(A)|$
  steps. Computing the most permissive controller for $A^i_p$ avoiding
  $\bad^\otimes_i$ can be done in linear time in the size of the
  region graph of $A^i_p$. As $RG(A^i)$ is a sub-graph of $RG(A^1)$,
  $RG(A_p^i)$ is a sub-graph of $RG(A_p^1)$. So we have to solve at
  most $|RG(A)|$ safety games of sizes at most $|RG(A_p^1)|$. As $A^1$
  is a sub-graph of $A^0_p=A^0\times_{\Sigma_l}A^0_2$, $|RG(A^1)| \leq
  |RG(A)|^2$. And as $A^1_p=A^1\times_{\Sigma_l}A^1_2$, $|RG(A_p^1)|
  \leq |RG(A)|^3$. So, $C^*$ can be computed in
  $O(|RG(A)|.|RG(A_p^1)|) = O(|RG(A)|^4) = O(2^{4.|A|})$. 
\end{IEEEproof}
\begin{theorem}\label{th-snni-cp-csp}
 For $\TAldet$, the SNNI-CP and SNNI-CSP are  EXPTIME-complete.
\end{theorem}
For the special case of finite automata we even have:
\begin{lemma}\label{lem-thm-converge-fa}
  For finite automata $C^*=C^{2}$.
\end{lemma}
 \begin{IEEEproof}
   We know that $\lang(C^2(A)\backslash \Sigma_h) \subseteq
   \lang(C^1(A)\backslash \Sigma_h)$.  Suppose that $\exists w$ \st $w
   \in \lang(C^1(A)\backslash \Sigma_h)$ and $w \not \in
   \lang(C^2(A)\backslash \Sigma_h)$ ($w$ cannot not be the empty
   word). We can assume that $w=u.l$ with $u \in \Sigma_l^*$, $l \in
   \Sigma_l \cap \Sigma_c$ and $u \in \lang(C^1(A)\backslash
   \Sigma_h)$ and $u.l \not\in \lang(C^2(A)\backslash \Sigma_h)$ ($l$
   is the first letter which witnesses the non membership property).
   If $l$ had to be pruned in the computation of $C^2$, it is because
   there is a word $u.l.m$ with $m \in \Sigma_u^*$ \st
   $\proj{\Sigma_l}(u.l.m) \in \lang(C^1(A)/\Sigma_h)$ but
   $\proj{\Sigma_l}(u.l.m) \not\in \lang(C^1(A) \backslash
   \Sigma_h)$. But by definition of $C^1$, $\lang(C^1(A)/\Sigma_h)
   \subseteq \lang(A\backslash \Sigma_h)$ (Lemma~\ref{lem-thm-2}) and
   thus $\proj{\Sigma_l}(u.l.m) \in \lang(A\backslash \Sigma_h)$.  As
   $u.l \in \Sigma_l^*$, $\proj{\Sigma_l}(u.l.m) =
   u.l.\proj{\Sigma_l}(m)$ and $\proj{\Sigma_l}(m) \in \Sigma_u^*$.
   Since $u.l \in \lang(C^1(A)\backslash \Sigma_h)$ and
   $\proj{\Sigma_l}(m) \in \Sigma_u^*$, we have
   $u.l.\proj{\Sigma_l}(m) \in \lang(C^1(A)\backslash \Sigma_h)$ which
   is a contradiction. Thus $\lang(C^2(A)\backslash \Sigma_h) =
   \lang(C^1(A)\backslash \Sigma_h)$ which is our stopping condition by
   lemma~\ref{lemma-snniifequal} and thus $C^*=C^2$.  
 \end{IEEEproof}
It follows that:
\begin{theorem}\label{th-finitedta-snnicsp}
  For a finite automaton $A$ in $\TAldet$ (i.e. such that $A \backslash \Sigma_h$ is
  deterministic), the SNNI-CSP is PSPACE-complete.
\end{theorem}
As untimed automata can always be determinized, we can extend our
algorithm to untimed automata when $A \backslash \Sigma_h$
non-deterministic. It suffices to determinize $A_2^i, i=1,2$:
\begin{theorem} \label{th-finite-snnicsp}
  For a finite automaton $A$ such that $A \backslash \Sigma_h$ is non
  deterministic, the SNNI-CSP can be solved in EXPTIME.
\end{theorem}

\begin{proposition}
  There is a family of finite automata $(A_i)_{i \geq 0}$ such that: $(i)$ there is
  a most permissive controller $D^*_i$ \st $D^*_i(A_i)$ is SNNI and
  $(ii)$ the memory required by $D^*_i$ is exponential in the size of
  $A_i$.
\end{proposition}
\begin{IEEEproof}

  Let $A$ be a finite automaton over the alphabet $\Sigma$.  Define
  the automaton $A'$ as given by Figure~\ref{fig-SNNI-det1}.  Assume
  the automaton $B$ is the sub-automaton of $A'$ with initial state
  $q'_0$.  We take $\Sigma_h=\{h\}=\Sigma_u$ and
  $\Sigma_l=\Sigma=\Sigma_c$. The most permissive controller $D$ \st
  $D(A')$ is SNNI generates the largest sub-language of $\lang(A')$
  \st $\lang(A' \backslash \Sigma_h)=\lang(A' / \Sigma_h)$ and thus it
  generates $\lang(A)=\lang(A' \backslash \Sigma_h)$.

  The controller $D$ is memoryless on $A' \backslash \Sigma_h$ as
  emphasized in Lemma~\ref{lem-thm-2}. It needs finite memory on the
  remaining part \ie on $B$.  
The controller $D$ on $B$ gives for each run a set of events of
$\Sigma$ that can be enabled: $D(q_0 \xrightarrow{\ h \ } q'_0
\xrightarrow{\ w \ } q'_0)=X$ with $w \in \Sigma^*$ and $X \subseteq
\Sigma_l$.As $B$ is deterministic, $D$ needs only the knowledge of $w$
and we can write $D(hw)$ ignoring the states of $A'$. For $B$ we can
even write $D(w)$ instead of $D(hw)$. Define the equivalence relation
$\equiv$ on $\Sigma^*$ by: $w \equiv w'$ if $D(w)=D(w')$.  Denote the
class of a word $w$ by $[w]$.  Because $D$ is memory bounded,
$\Sigma^*_{/ \equiv}$ is of finite index which is exactly the memory
needed by $D$.

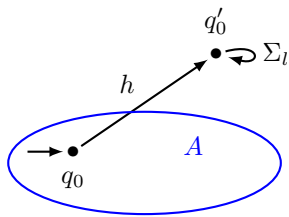
\begin{figure}  \centering
  \begin{tikzpicture}[thick,node distance=1.3cm and 1.9cm,every
    state/.style={minimum size=.1cm,draw=none,inner sep=0}]%
    \node[state,initial] (q_0) [label=below:{$q_0$}] {$\bullet$}; 
    \node[state] (q_1) [above right=of q_0,label=above:{$q'_0$}] {$\bullet$};
    \node (q_2) [right=of q_0,xshift=.4cm,yshift=-.5cm] {};
    \node (q_3) [above=of q_2,yshift=-.6cm] {};

    \path[->] (q_0)  edge  node  {$h$} (q_1)
              (q_1) edge [loop right] node  {$\Sigma_l$} (q_1);

\node[draw=blue,inner sep=0pt,xshift=-.2cm,thick,ellipse,fit=(q_0) (q_2) (q_3)] {}; 
\node (x) [right=of q_0,shift={(-.3cm,.07cm)}] {\textcolor{blue}{$A$}};
  \end{tikzpicture}
   \caption{Automaton $B$}
   \label{fig-SNNI-det1}
\end{figure}
Thus we can define an automaton $D_{/ \equiv}=(M,$
$m_0,\Sigma,\rightarrow)$ by: $M=\{[w] \ | \ w \in \Sigma^*\}$,
$m_0=[\varepsilon]$, and $[w] \xrightarrow{\ a \ } [wa]$ for $a \in
D(hw)$.  $D_{/ \equiv}$ is an automaton which accepts $\lang(A)$ (and
it is isomorphic to $D(B)$) and the size of which is the size of $D$
because $B$ has only one state.  This automaton is deterministic and
thus $D_{/ \equiv}$ is also deterministic and accepts $\lang(A)$.
There is a family $(A_i)_{i \geq 0}$ of non-deterministic finite automata, such
that the deterministic and language-equivalent automaton of each $A_i$
requires at least exponential size. For each of these $A_i$ we
construct the controller $D_{/ \equiv}^i$ as described before, and
this controller must have at least an exponential size (\wrt to
$A_i$). This proves the EXPTIME lower bound.  
\end{IEEEproof}

In this section we have studied the strong non-deterministic
non-interference control problem (SNNI-CP) and control synthesis
problem (SNNI-CSP) in the timed setting. The main results we have
obtained are: (1) the SNNI-CP can be solved if $A \backslash
\Sigma_h$ can be determinized and is undecidable otherwise; (2) the
SNNI-CSP can be solved by solving a finite sequence of safety games if
$A \backslash \Sigma_h$ can be determinized.
We have provided an optimal algorithm to solve the SNNI-CP and CSP
in this case (although we have not proved a completeness result).
\begin{table}[hbtp]
\begin{center}
\begin{tabular}{||l||c|c||c|c||}\hline\hline
   & \multicolumn{2}{c||}{$A$ Timed Automaton} & \multicolumn{2}{c||}{$A$ Finite Automaton}\\%\cline{2-5}
  & {$A \backslash \Sigma_h$ Non-Det.~} & {$A \backslash \Sigma_h$ Det.~} & {$A \backslash \Sigma_h$ Non-Det.~} & {$A \backslash \Sigma_h$ Det.~}  \\\hline\hline 
%  {SNNI-VP} & {undecidable} & {PSPACE-C} & {PSPACE-C} & PTIME \\ \hline
  {SNNI-CP} & {undecidable} (Theorem~\ref{thm-undecidable}) & {EXPTIME-C} (Theorem~\ref{th-snni-cp-csp}) & {PSPACE-C} (Theorem~\ref{thm-0}) & PTIME (Corollary~\ref{cor-SNNI-CP-fdta}) \\ \hline
  {SNNI-CSP} & {undecidable} (Theorem~\ref{thm-undecidable}) & {EXPTIME-C} (Theorem~\ref{th-snni-cp-csp}) & {EXPTIME} (Theorem~\ref{th-finite-snnicsp}) & PSPACE-C  (Theorem~\ref{th-finitedta-snnicsp}) \\ \hline\hline
\end{tabular}
\end{center}
\caption{Summary of the Results for SNNI-CP and SNNI-CSP}
\label{tab-summary}
\end{table}

The summary of the results is given in Table~\ref{tab-summary}.  
%For non-deterministic FA, we have proved that EXPTIME is a lower bound.

\section{BSNNI and CSNNI Control Problems}
\label{sec-bcsnni-csp}

In this section, we will show that for more restrictive
non-interference properties (CSNNI and BSNNI) the control problem
presents a major drawback: in the general case, there is no most
permissive controller.

The CSNNI-\emph{Control Problem} CSNNI-CP (respectively
BSNNI-\emph{Control Problem} BSNNI-CP) we are interested in is the
following:
\begin{equation}\label{eq-bsni-rp} \tag{CSNNI-CP, BSNNI-CP}
  \textit{Is there a controller $C$  s.t. $C(A)$ is CSNNI (respectively BSNNI) ?}
\end{equation}

The CSNNI-\emph{Controller Synthesis Problem} CSNNI-CSP (respectively
BSNNI-\emph{Controller Synthesis Problem} BSNNI-CSP) asks to compute a
witness when the answer to the CSNNI-CP (respectively BSNNI-CSP) is
``yes''.

%We will now prove that, for finite automata, the CSNNI-CP can be solve in a manner similar to that of SNNI-CP:

\subsection{CSNNI-CP and CSNNI-CSP}

\begin{theorem}\label{thm-snnicpuntimed}
  For finite automata the CSNNI-CP is in PTIME.
\end{theorem}

\begin{IEEEproof}

  Let $\M$, be a finite automaton, we show that there exists a
  controller $C$ such that $C(\M)$ is CSNNI if and only if $\M
  \backslash \Sigma_c$ is CSNNI.

  The \emph{if} direction is obvious: the controller $C_\forall$ that
  prevents any controllable action from occurring is defined by:
  $C_\forall(\rho)=\emptyset$, $\forall \rho \in \runs(\M)$. It is
  easy to see that $C_\forall(\M)$ is isomorphic to $\M \backslash
  \Sigma_c$ and thus bisimilar.

  This \emph{only if} direction is proved as follows: let $A_1$ and
  $A_2$ be two finite automata over alphabet $\Sigma^\varepsilon$ such
  that $A_1$ weakly simulates $A_2$.  Consider $A'_1 = A_1 \backslash
  \{e\}$ and $A'_2 = A_2 \backslash \{e\}$ for $e \in \Sigma$. Clearly,
$A'_1$ simulates $A'_2$ (by definition of the simulation relation).

Therefore, if there exists $C$ \st $C(A)$ is CSNNI, then so is
$C(A)\backslash \Sigma'$ for any $\Sigma' \subseteq \Sigma$.  It
follows that $C(A) \backslash \Sigma_c$ must be CSNNI.

The CSNNI-CP reduces to the CSNNI-VP which is PTIME for finite
automata.

\end{IEEEproof}

\begin{theorem}\label{thm-snnicpfinite}
  For the class of deterministic finite automata, the CSNNI-CSP is
  PSPACE-complete.
\end{theorem}
\begin{IEEEproof}
  By Lemma~\ref{lem-csnnisnnidet}, for deterministic automata, SNNI is
  equivalent to CSNNI. Hence the CSNNI-CSP is equivalent to the
  SNNI-CSP which is PSPACE-complete by
  Theorem~\ref{th-finitedta-snnicsp}.
\end{IEEEproof}

In the timed setting, the previous reduction to a verification problem
cannot be applied as illustrated by the following
% this algorithm does not work to resolve the CSNNI-CP as it it
% illustrated by the
example~\ref{ex-snnicpuntimedproof}.

\begin{example}
%[Counterexample]
\label{ex-snnicpuntimedproof}

  Let $\M$ be the deterministic timed automaton given in
  figure~\ref{fig-snnicpuntimedproof1} with $\Sigma_l=\{\ell_1,
  \ell_2\}$, $\Sigma_h=\{h\}$ and $\Sigma_c=\{\ell_1\} $. $\M
  \backslash \Sigma_c$ is neither CSNNI nor SNNI (here SNNI and CSNNI
  are equivalent since $\M$ is deterministic). However there exists a
  controller $C$ such that $C(\M)$ is both CSNNI and SNNI. $C(\M)$ can
  be given by the timed automaton given in
  figure~\ref{fig-snnicpuntimedproof2}.

 	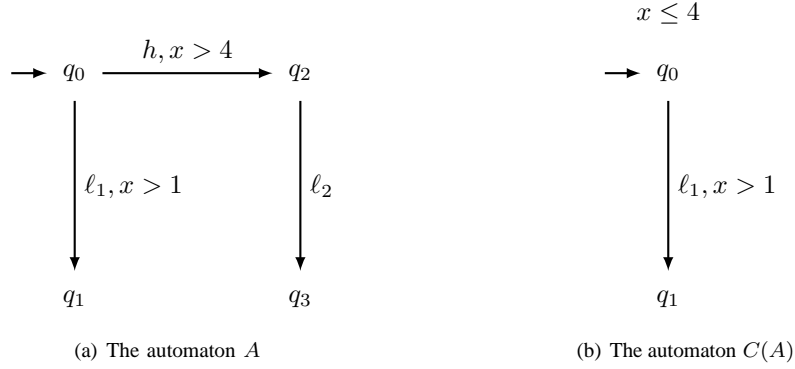
\begin{figure}
\begin{center}

		\subfigure[The automaton $\M$]{\label{fig-snnicpuntimedproof1}
\begin{tikzpicture}[thick,node distance=3cm,initial text=,auto]% 
    \node[state,initial] (a) {$q_0$}; 
    \node[state] (b) [below of=a] {$q_1$};
    \node[state] (c) [right of=a] {$q_2$};
    \node[state] (d) [below of=c] {$q_3$};
    \path[->] (a)  edge node  {$\ell_1,x > 1$} (b);
    \path[->] (a)  edge node  {$h,x >4$} (c);
    \path[->] (c)  edge node  {$\ell_2$} (d);
 \end{tikzpicture}
		}\hspace{3cm}%
		\subfigure[The automaton $C(\M)$]{\label{fig-snnicpuntimedproof2}
		\begin{tikzpicture}[thick,node distance=3cm,initial text=,auto]% 
		    \node[state,initial] (a) {$q_0$}; 
		    \node[state] (b) [below of=a] {$q_1$};
                    \node (y) [above of =a,shift={(0cm,-2.2cm)}] {$x\leq 4$};
		     \path[->] (a)  edge node  {$\ell_1,x > 1$} (b);
		\end{tikzpicture}
		}
		\caption{Counterexample of theorem~\ref{thm-snnicpuntimed} in timed setting}
		\label{fig-snnicpuntimedproof12}
\end{center}
	\end{figure}
 
\end{example}

However for the timed automata in $\TAldet$, thanks to
Lemma~\ref{lem-csnnisnnidet} and Theorems~\ref{th-snni-cp-csp}
and~\ref{th-finitedta-snnicsp}, we have:

\begin{theorem}\label{thm-snnicpcspdta}
  For timed automata in $\TAldet$, the CSNNI-CP and CSNNI-CSP are
  EXPTIME-complete.
%For the class of finite automata in $\TAldet$,  the CSNNI-CSP is PSPACE-complete.
\end{theorem}
\begin{IEEEproof}
  By Lemma~\ref{lem-csnnisnnidet} the CSNNI-CP/CSNNI-CSP is equivalent
  to the SNNI-CP/SNNI-CSP for $\TAldet$ and by
  Theorem~\ref{th-snni-cp-csp}, it follows that CSNNI-CP and CSNNI-CSP
  are EXPTIME-complete.
\end{IEEEproof}

Moreover, for $\TAldet$, thanks to the algorithm of
section~\ref{sec-results} there always exists a most permissive
controller for CSNNI.
However we will now show that there is a  non-deterministic finite automaton
\st there is no most permissive controller ensuring CSNNI.
% there not
%always exists a most permissive controller ensuring CSNNI.

\begin{proposition} \label{prop-CSNNI-nmp} There is no most permissive
  controller ensuring CSNNI for the finite automaton $A \not \in \TAldet $  of
  figure~\ref{fig-snnipascsnni1} (i.e. such that $A \backslash \Sigma_h$ is
  non deterministic) with $\Sigma_h=\{h\}$,
  $\Sigma_l=\{\ell_1,\ell_2,\ell_3\}$ and
  $\Sigma_c=\{\ell_2,\ell_3\}$.
\end{proposition}

\begin{IEEEproof}

  Let $\M_c$ be the finite automaton of figure~\ref{fig-snnipascsnni1}
  with $\Sigma_h=\{h\}$, $\Sigma_l=\{\ell_1,\ell_2,\ell_3\}$ and
  $\Sigma_c=\{\ell_2,\ell_3\}$. $\M_c \not \in \TAldet$ since $\M_c \backslash \Sigma_h$ is
  non-deterministic. This automaton is not CSNNI. The
  controllers $C_1$ and $C_2$ of figure~\ref{fig-csnnicontrole23} make
  the system CSNNI. However $(C_1\cup C_2)(\M_c)=\M_c$ is not CSNNI
  and, by construction is the only possible controller more permissive
  than $C_1$ and $C_2$. Therefore, there is no most permissive
  controller ensuring CSNNI for $\M_c$ with $\Sigma_c$.
 
 	\begin{figure}
		\subfigure[Automaton $C_1(\M_c)$]{\label{fig-csnnicontrole2}
		\begin{tikzpicture}[node distance=1.5cm,thick,initial text=,auto,scale=0.7]
			%\gasset{ExtNL=y,NLdist=1,NLangle=90}
			\node[state,initial] (q0) at (0,0) {$q_0$};
			\node(q) [below of=q0] {};
			\node[state] (q1) [left of=q] {$q_1$};
			\node[state] (q2) [right of=q] {$q_2$};
			\node[state] (q3) [below of=q1] {$q_3$};
			\node[state] (q4) [below of=q2] {$q_4$};
			\node(qq) [right of=q0] {};
			\node(qqq) [right of=qq] {};
			\node[state] (q5) [right of=qqq] {$q_5$};
			\node[state] (q6) [below of=q5] {$q_6$}; 
			\node(qp) [below of=q6] {}; 
			\node[state] (q7) [left of=qp] {$q_7$};  

			%\nodelabel[NLangle=90,NLdist=6](a){$[x_1 \leq 4]$}

			\path[->] (q0) edge node {$\ell_1$} (q1);
			\path[->] (q0) edge node {$\ell_1$} (q2);
			\path[->] (q1) edge node {$\ell_2$} (q3);
			\path[->] (q2) edge node {$\ell_3$} (q4);
			\path[->] (q0) edge node {$h$} (q5);
			\path[->] (q5) edge node {$\ell_1$} (q6);
			\path[->] (q6) edge node {$\ell_2$} (q7);
		\end{tikzpicture}
		}\hspace{15pt}
		\subfigure[Automaton $C_2(\M_c)$]{\label{fig-csnnicontrole3}
		\begin{tikzpicture}[node distance=1.5cm,thick,initial text=,auto,scale=0.7]
			%\gasset{ExtNL=y,NLdist=1,NLangle=90}
			\node[state,initial] (q0) at (0,0) {$q_0$};
			\node(q) [below of=q0] {};
			\node[state] (q1) [left of=q] {$q_1$};
			\node[state] (q2) [right of=q] {$q_2$};
			\node[state] (q3) [below of=q1] {$q_3$};
			\node[state] (q4) [below of=q2] {$q_4$};
			\node(qq) [right of=q0] {};
			\node(qqq) [right of=qq] {};
			\node[state] (q5) [right of=qqq] {$q_5$};
			\node[state] (q6) [below of=q5] {$q_6$}; 
			\node(qp) [below of=q6] {};  
			\node[state] (q8) [right of=qp]  {$q_8$};

			%\nodelabel[NLangle=90,NLdist=6](a){$[x_1 \leq 4]$}

			\path[->] (q0) edge node {$\ell_1$} (q1);
			\path[->] (q0) edge node {$\ell_1$} (q2);
			\path[->] (q1) edge node {$\ell_2$} (q3);
			\path[->] (q2) edge node {$\ell_3$} (q4);
			\path[->] (q0) edge node {$h$} (q5);
			\path[->] (q5) edge node {$\ell_1$} (q6);
			\path[->] (q6) edge node {$\ell_3$} (q8);
		\end{tikzpicture}
		}
		\caption{Automata $C_1(\M_c)$ and $C_2(\M_c)$}
		\label{fig-csnnicontrole23}
	\end{figure}
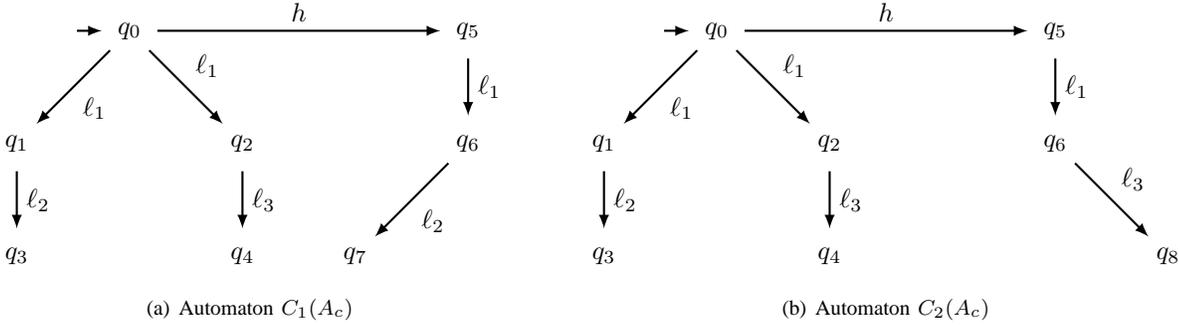

\end{IEEEproof}

\subsection{BSNNI-CP and BSNNI-CSP}
%\FC{as BSNNI implies CSNNI the example below looks redundant as we
%  have already showed this for CSNNI. Correct ? Or do we need a new
%  example and then we should say why. Sa me holds for proposition 5 below.} 
%%
%\OR{Pour l'exemple, je ne sais pas trop si c'est \`evident, en tout
%  cas, le meme exemple ne marche pas puisque l'exemple pour la SNNI et
%  CSNNI n'est pas BSNNI. Pour la proposition 5 : le resultat est plus
%  fort car il n'y a pas forcement de controller le plus permissif meme
%  si l'automate fini est deterministe alors que pour CSNNI, quand il
%  est deterministe ou a une solution. }
%
%\FC{OK ! j'ajoute un epetite phrase avant prop 5 alors.}

We first show by example~\ref{ex-bsnnicp} that even if there exists a
controller for a finite automaton $\M$ and a controllable alphabet
$\Sigma_c$ ensuring BSNNI (i.e. the answer to BSNNI-CP is $\true$), it
is possible to have $\M \backslash \Sigma_c$ not BSNNI.

\begin{example}\label{ex-bsnnicp} 
  Let $\M_i$ be the finite automaton of figure~\ref{fig-bsnnicp} with
  $\Sigma_h=\{h_1,h_2\}$ et $\Sigma_l=\{\ell\}$. This automaton is
  BSNNI, then the answer to BSNNI-CP is $\true$ for all
  $\Sigma_c$. However, for $\Sigma_c=\{h_2\}$, the automaton $\M_i
  \backslash \Sigma_c=\M_e$ %of figure~\ref{fig-csnnipasbsnni1}
  is not BSNNI.

 	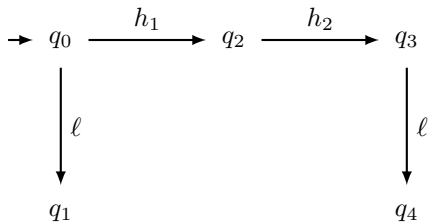
\begin{figure}
\begin{center}
 
		\begin{tikzpicture}[node distance=2.3cm,thick,initial text=,auto,scale=0.7]
			%\gasset{ExtNL=y,NLdist=1,NLangle=90}
			\node[state,initial] (q0) at (0,0) {$q_0$};
			\node[state] (q1) [below of=q0] {$q_1$};
			\node[state] (q2) [right of=q0] {$q_2$};   
			\node[state] (q3) [right of=q2] {$q_3$};
			\node[state] (q4) [below of=q3] {$q_4$};
			%\nodelabel[NLangle=90,NLdist=6](a){$[x_1 \leq 4]$}

			\path[->] (q0) edge node {$\ell$} (q1);
			\path[->] (q0) edge node {$h_1$} (q2);
			\path[->] (q2) edge node {$h_2$} (q3);
			\path[->] (q3) edge node {$\ell$} (q4);
		\end{tikzpicture}
		\caption{The automaton $\M_i$}
		\label{fig-bsnnicp}
\end{center}
	\end{figure}
\end{example}

We will now prove that for deterministic finite automaton 
%(and of course we can extend the result to timed automata) even deterministic, there not
%always exists a most permissive controller ensuring BSNNI.
there is not always a most permissive controller that enforces BSNNI.
This result is in contrast with CSNNI where a most permissive controller
always exists for  $\TAldet$.

\begin{proposition} \label{prop-BSNNI-nmp} There is no most permissive
  controller ensuring BSNNI for the deterministic finite automaton of
  figure~\ref{fig-csnnipasbsnni1} with $\Sigma_h=\{h\}$,
  $\Sigma_l=\{\ell\}$ and $\Sigma_c=\{\ell,h\}$.
\end{proposition}

\begin{IEEEproof}

  Let $\M_e$ be the deterministic finite automaton of
  figure~\ref{fig-csnnipasbsnni1} with $\Sigma_h=\{h\}$,
  $\Sigma_l=\{\ell\}$ and $\Sigma_c=\{\ell,h\}$. This automaton is not
  BSNNI. The controllers $C_1$ and $C_2$ of
  figure~\ref{fig-bsnnicontrole23} make the system BSNNI.  However,
  $(C_1\cup C_2)(\M_e)=\M_e$ is not BSNNI and, by construction is the
  only possible controller more permissive than $C_1$ and
  $C_2$. Therefore, there is no most permissive controller ensuring
  BSNNI for $\M_e$ with $\Sigma_c$.

 	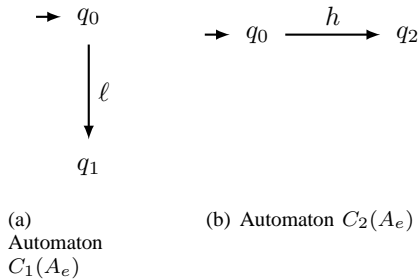
\begin{figure}
	  \begin{center}
		\subfigure[Automaton $C_1(\M_e)$]{\label{fig-bsnnicontrole2}
		\begin{tikzpicture}[node distance=2cm,thick,initial text=,auto,scale=0.7]
			%\gasset{ExtNL=y,NLdist=1,NLangle=90}
			\node[state,initial] (q0) at (0,0) {$q_0$};
			\node[state] (q1) [below of=q0] {$q_1$};   
			%\nodelabel[NLangle=90,NLdist=6](a){$[x_1 \leq 4]$}

			\path[->] (q0) edge node {$\ell$} (q1);
		\end{tikzpicture}
		}\hspace{15pt}
		\subfigure[Automaton  $C_2(\M_e)$]{\label{fig-bsnnicontrole3}
		\begin{tikzpicture}[node distance=2cm,thick,initial text=,auto,scale=0.7]
			%\gasset{ExtNL=y,NLdist=1,NLangle=90}
			\node[state,initial] (q0) at (0,0) {$q_0$};
			\node[state] (q2) [right of=q0] {$q_2$};  
			\node (q1) [below of=q0] {};  
			%\nodelabel[NLangle=90,NLdist=6](a){$[x_1 \leq 4]$}
			
			\path[->] (q0) edge node {$h$} (q2);
		\end{tikzpicture}
		}
		\caption{Automata $C_1(\M_e)$ and $C_2(\M_e)$}
		\label{fig-bsnnicontrole23}
	  \end{center}
	\end{figure}
\end{IEEEproof}

\begin{table}[hbtp]
\begin{center}
\begin{tabular}{||l||c|c||c|c||}\hline\hline
   & \multicolumn{2}{c||}{$A$ Timed Automaton} & \multicolumn{2}{c||}{$A$ Finite Automaton}\\%\cline{2-5}
  & {$A \backslash \Sigma_h$ Non-Det.~} & {$A \backslash \Sigma_h$ Det.~} & {$A \backslash \Sigma_h$ Non-Det.~} & {$A \backslash \Sigma_h$ Det.~}  \\\hline\hline 
  {CSNNI-CP} & {open} & {EXPTIME-C}  (Theorem~\ref{thm-snnicpcspdta}) & PTIME (Theorem~\ref{thm-snnicpuntimed}) & PTIME (Theorem~\ref{thm-snnicpuntimed}) \\ \hline
  {CSNNI-CSP} & {\nmp}~(Proposition~\ref{prop-CSNNI-nmp}) & {EXPTIME-C} (Theorem~\ref{thm-snnicpcspdta}) & {\nmp}~(Proposition~ \ref{prop-CSNNI-nmp}) & PSPACE-C (Theorem~\ref{thm-snnicpfinite}) \\ \hline
  {BSNNI-CSP} & {\nmp}~(Proposition~\ref{prop-BSNNI-nmp}) & \nmp~(Proposition~\ref{prop-BSNNI-nmp}) & \nmp~(Proposition~\ref{prop-BSNNI-nmp}) & \nmp~(Proposition~\ref{prop-BSNNI-nmp}) \\ \hline\hline
\end{tabular}
\end{center}
\hspace{2 cm}* NMPC means that there not always exists a most permissive controller.
\caption{Summary of the Results for CSNNI and BSNNI Control Problems}
\label{tab-summary-bc-snni}
\end{table}

The summary of the results for CSNNI and BSNNI Control Problems is given in Table~\ref{tab-summary-bc-snni}.

\section{Conclusion and Future Work}
\label{sec-conclu}

In this paper we have studied the strong non-deterministic
non-interference control problem and control synthesis problem in the
timed setting. The main results we have obtained are: (1) the SNNI-CP
can be solved if $A \backslash \Sigma_h$ can be determinized and is
undecidable otherwise; (2) the SNNI-CSP can be solved by solving a
finite sequence of safety games if $A \backslash \Sigma_h$ can be
determinized; (3) there is not always a least restrictive (most
permissive) controller for (bi)simulation based non-interference even
for untimed finite automata. However, there is a most permissive
controller for CSNNI if $A \backslash \Sigma_h$ is deterministic and
CSNNI-CP and CSNNI-CSP are EXPTIME-complete in this case in the timed
setting.

The summary of the results is given in Tables~\ref{tab-res-snni-vp}
and~\ref{tab-res-bcsnni-vp} for the verification problems and
Tables~\ref{tab-summary} and~\ref{tab-summary-bc-snni} for the control
problems.

\noindent Our future work will focus on the CSNNI-CP (and BSNNI-CP)
as 
%to solve CSNNI-CP and
%  BSNNI-CP 
even when there is no most permissive controller it is interesting to
find one. Another future direction will consist in determining
conditions under which a least restrictive controller exists for the
BSNNI-CSP.

%\bibliographystyle{elsarticle-num}
%\bibliography{biblio}

\end{document}